\documentclass[final,3p,times]{elsarticle}
\usepackage{graphicx}
\usepackage{txfonts}
\usepackage[ansinew]{inputenc}
\usepackage[T1]{fontenc}
\usepackage{bm}
\usepackage{amssymb}
\newcommand{\sty}{\scriptstyle}
\newcommand{\ssty}{\scriptscriptstyle}
\newcommand{\tsty}{\textstyle}
\newcommand{\be}{\begin{equation}}
\newcommand{\ee}{\end{equation}}
\newcommand{\lb}[1]{\label{#1}}
\newcommand{\dl}{d_{\ssty L}}
\newcommand{\da}{d_{\ssty A}}
\newcommand{\dg}{d_{\ssty G}}
\newcommand{\dz}{d_z}
\newcommand{\gaml}{\gamma_{\ssty L}}

\newcommand{\gamg}{\gamma_{\ssty G}}
\newcommand{\gamz}{\gamma_{\ssty z}}
\newcommand{\df}{\mathrm{d}}
\newcommand{\obs}[1]{[#1]_{{\tsty {\ssty \rm obs}}}}
\newcommand{\bigobs}[1]{\left[#1\right]_{{\tsty {\ssty \rm obs}}}}
\newcommand{\Mbar}{\bar{M}}
\newcommand{\etal}{et al.\ }
\journal{Physica A}
\begin{document}
\begin{frontmatter}
%% Title, authors and addresses
%% use the tnoteref command within \title for footnotes;
%% use the tnotetext command for the associated footnote;
%% use the fnref command within \author or \address for footnotes;
%% use the fntext command for the associated footnote;
%% use the corref command within \author for corresponding author footnotes;
%% use the cortext command for the associated footnote;
%% use the ead command for the email address,
%% and the form \ead[url] for the home page:
%%
%% \title{Title\tnoteref{label1}}
%% \tnotetext[label1]{}
%% \author{Name\corref{cor1}\fnref{label2}}
%% \ead{email address}
%% \ead[url]{home page}
%% \fntext[label2]{}
%% \cortext[cor1]{}
%% \address{Address\fnref{label3}}
%% \fntext[label3]{}
%% use optional labels to link authors explicitly to addresses:
%% \author[label1,label2]{<author name>}
%% \address[label1]{<address>}
%% \address[label2]{<address>}
\title{Fractal analysis of the galaxy distribution in the redshift range $0.45 \leq z \leq 5.0$}
\author[label1]{G.\ Conde-Saavedra}
\ead{gazzurra@gmail.com}
\author[label1]{A.\ Iribarrem}
\ead{airibarrem@gmail.com}
\author[label2]{Marcelo B.\ Ribeiro\corref{cor1}}
\ead{mbr@if.ufrj.br}
\cortext[cor1]{Corresponding author}
\address[label1]{Observat\'orio do Valongo, Universidade Federal do Rio de Janeiro, Brazil}
\address[label2]{Instituto de F\'{\i}sica, Universidade Federal do Rio de Janeiro, Brazil}
\begin{abstract}
This paper performs a fractal analysis of the galaxy distribution
and presents evidence that it can be described as a fractal system
within the redshift range of the FORS Deep Field (FDF) galaxy survey
data. The fractal dimension $D$ was derived by means of the galaxy
number densities calculated by Iribarrem et al.\ (2012a) %, A\&A, 539, A112)
using the FDF luminosity function parameters and absolute magnitudes
obtained by Gabasch et al.\ (2004, 2006) in the spatially homogeneous
standard cosmological model with $\Omega_{m_0}=0.3$, $\Omega_{\Lambda_0}
=0.7$ and $H_0=70 \; \mbox{km} \; {\mbox{s}}^{-1} \; {\mbox{Mpc}}^{-1}$.
Under the supposition that the galaxy distribution forms a fractal
system, the ratio between the differential and integral number densities
$\gamma$ and $\gamma^\ast$ obtained from the red and blue FDF galaxies
provides a direct method to estimate $D$ and implies that $\gamma$ and
$\gamma^\ast$ vary as power-laws with the cosmological distances,
feature which provides a second method for calculating $D$. The
luminosity distance $\dl \,$, galaxy area distance $\dg$ and redshift
distance $\dz$ were plotted against their respective number densities
to calculate $D$ by linear fitting. It was found that the FDF galaxy
distribution is better characterized by two single fractal dimensions
at successive distance ranges, that is, two scaling ranges in the
fractal dimension. Two straight lines were fitted to the data, whose
slopes change at $z \approx 1.3$ or $z \approx 1.9$ depending on the
chosen cosmological distance. The average fractal dimension calculated
using $\gamma^\ast$ changes from
$\langle D \rangle=1.4^{\ssty +0.7}_{\ssty -0.6}$ to
$\langle D \rangle=0.5^{\ssty +1.2}_{\ssty -0.4}$ for all galaxies.
Besides, $D$ evolves with $z \,$, decreasing
as the redshift increases. Small values of $D$ at high $z$ mean that
in the past galaxies and galaxy clusters were distributed much more
sparsely and the large-scale structure of the universe was then
possibly dominated by voids.
The results of Iribarrem \etal (2014) indicating that similar fractal
features having $\langle D \rangle =0.6 \pm 0.1$ can be found in the
100 $\mu$m and 160 $\mu$m passbands of the far-infrared sources of
the Herschel/PACS evolutionary probe (PEP) at $1.5 \lesssim z \lesssim
3.2$ are also mentioned.
\end{abstract}
\end{frontmatter}
% \linenumbers
%% main text
%
\begin{flushright}
\textit{``Inside of every large problem is a small problem that would
simply resolve the big\\problem, but which will not be discovered
because everyone is working on the large problem.''}\\Hoare's Law of
Large Problems
\end{flushright}
\section{Introduction}\lb{intro}

Fractal analysis of the galaxy distribution consists
of using the standard techniques of fractal geometry to verify
whether or not a given galaxy distribution has fractal properties
and calculating its key feature, the \textit{fractal dimension}
$D$, with data gathered from the distribution. The fractal dimension
quantifies how ``broken'', or irregular, is the distribution, that is,
how far a distribution departs from regularity. Hence, in the context
of the galaxy distribution the fractal dimension is simply a measure
of the possible degree of inhomogeneity in the distribution,
which can be viewed as a measure of galactic clustering sparsity
or, complementarily, the dominance of voids in the large-scale
structure of the Universe. Values of $D$ smaller than the
corresponding topological dimension where the fractal system
is embedded mean a more irregular pattern (Mandelbrot 1983).
Systems having three dimensional topology such as the galaxy
distribution are regular, or homogeneous, if $D=3$. Accordingly,
increasing irregularities, or inhomogeneities, in the
distribution corresponds to decreasing values of the fractal
dimension, that is, $D<3$ (Ribeiro and Miguelote 1998; Sylos
Labini et al.\ 1998).

Fractal distributions described by one fractal dimension are called
\textit{single fractals} and form the simplest fractal systems. This
is, of course, a simplification, but a useful one as a first approach
for describing complex distributions or for analyzing simple systems
(Ribeiro and Miguelote 1998). More complex distributions can exhibit
different values of $D$ at specific distance ranges defined in the
distribution, that is, different scaling ranges in the fractal
dimension so that $D=D(d)$ where $d$ is the distance. In such a case
there is a succession of single fractal systems. An even more complex
situation can occur if, for instance, quantities like mass or
luminosity range between very different values, i.e., if they possess
a distribution. Such variations require a generalization of the
fractal dimension in order to include the distribution and, hence,
the system is characterized by several fractal dimensions in the
same scaling range, that is, a whole spectrum of dimensions whose
maximum value corresponds to the single fractal dimension $D$ the
system would have if the studied quantity did not range. In such a
case the system is said to exhibit a \textit{multifractal} pattern
(Gabrielli et al.\ 2005).

Cosmological models which describe the galaxy distribution as a
fractal system are not new. Several of such studies can be found
in the literature, either in the context of Newtonian cosmology
(Pietronero 1987; Ribeiro and Miguelote 1998; Sylos Labini et al.\
1998; Abdalla et al.\ 1999; Gabrielli et al.\ 2005; Sylos Labini
2011; and references therein) or in models based on relativistic
cosmology (Ribeiro 1992ab, 1993, 1994, 2001ab, 2005; Abdalla et
al.\ 2001; Abdalla and Chirenti 2004; Mureika and Dyer 2004;
Mureika 2007; and references therein).  Fractal analyzes based on
Newtonian cosmology often perform detailed statistical tests on
empirical data, but they are usually limited to very small redshift
ranges where inhomogeneities predicted in standard relativistic
models are not detectable (Rangel Lemos and Ribeiro 2008). On the
other hand, relativistic cosmology fractal models had to cope with
fundamental conceptual problems like how to define fractality in
a curved spacetime and what is the meaning of observable homogeneity,
as opposed to spatial homogeneity (Ribeiro 1992ab, 1995, 2001b, 2005;
Rangel Lemos and Ribeiro 1998). Overcoming these difficulties led to 
mostly theoretical models with little or none empirical data analysis.

Studies not motivated by fractals sometimes provide, nevertheless,
data analyzes which can be used to quantitatively
measure statistical fractal properties in the galaxy distribution
because some of their results suggest a fractal pattern in this
distribution. This is the case of Albani et al.\ (2007; hereafter
A07) and, more recently, Iribarrem et al.\ (2012a, hereafter Ir12a;
see also Iribarrem et al.\ 2014)
who carried out relativistic analyzes of galaxy number densities
at high redshift ranges based on empirical data derived from the
galaxy luminosity function (LF). A07 used LF data from the CNOC2
galaxy redshift survey (Lin et al.\ 1999) in the range $0.1 \leq z
\leq 1.0$, whereas Ir12a (see also Iribarrem et al.\ 2012b) carried
out a similar analysis using LF data extracted from red and blue
galaxies belonging to the FORS Deep Field (FDF) galaxy redshift
survey (Gabasch et al.\ 2004, 2006; hereafter G04 and G06,
respectively) in the range $0.45 \leq z \leq 5.0$. Both studies
found evidence that at high redshifts the galaxy number densities
obtained from the LF scale as power-laws with the relativistic
distances. As it is well-known, distributions obeying
power-laws strongly suggest fractal behavior (Mandelbrot 1983).

{Measuring fractal properties from these data analyzes
became possible because the LF was computed using the
1/V$_{\mathrm{max}}$ method, which is a non-parametric estimator
that assumes a homogeneous distribution on average to correct for
the incompleteness caused by the flux limit of a survey. Subsequent
integration over absolute magnitudes produced the selection
function, which is essentially a galaxy number density that can be
transformed into other densities by using the relativistic cosmology
based framework developed by Ribeiro and Stoeger (2003), A07 and
Ir12a. Assuming that the LF parameters of G04 and G06 are not biased
by any other radial selection effects, then our interest was basically
focused on making sure that the number densities obtained by integrating
the LF were not biased by the integration limit. By using an absolute
magnitude cut based on the formal 50\% completeness limit of the
I-band, at which the galaxies were selected, Ir12a ended up with a
comoving number density that
% that is complete, in the sense of the original 1/V$_{max}$ method
% used in computing the LF, up to the volume defined by the
% magnitude cut. This yields a comoving density
corresponds to the brighter objects of the sample in separate
redshift bins. Such subsamples yield approximately the same number
density as the one computed straightforwardly from a volume-limited
sample. Together with correcting for flux limit incompleteness, done
in the building of the LF, keeping under control a possible bias
introduced by the redshift dependence of the integration limit is
an essential requirement for measuring fractal dimensions.}

The aim of this paper is to perform a relativistic analysis of the
results presented in Ir12a from a fractal perspective. We focused
on this work because it is based on the FDF survey, whose galaxies
were measured at several wavebands and out to deep redshift ranges.
Although this is a survey that scanned a limited sky area, its
redshift depth is the main feature that motivated this work, as
having measured galaxies up to $z=5.0$ the FDF survey is capable
of producing results that can indicate possible observable
inhomogeneities even when one uses the standard spatially homogeneous
Friedmann-{Lema\^{\i}tre}-Robertson-Walker (FLRW) cosmological
model and, hence, if a fractal approach for studying the galaxy
distribution is worth pursuing with other, less limited, deep
samples. In addition, since the results were obtained assuming
the FLRW cosmology with $\Omega_{m_{\ssty 0}} = 0.3$,
$\Omega_{\Lambda_{\ssty 0}} = 0.7$ and $H_0=70 \; \mbox{km} \;
{\mbox{s}}^{-1} \; {\mbox{Mpc}}^{-1}$, this means that the
cosmological principle will be valid in the whole analysis of
this paper. Actually, it has already been shown elsewhere that
there is no contradiction whatsoever between observational
fractals and the cosmological principle (Ribeiro 2001b, 2005;
Rangel Lemos and Ribeiro 2008).

The results show that the simplest fractal description of the
galaxy distribution in the redshift range $0.45 \leq z \leq 5.0$
needs two fractal dimensions associated to specific distance ranges
to describe the distribution. In other words, we found two scaling
ranges in the fractal dimension. The transition between these two
regions spans the range $z=1.3 - 1.9$. In the first region, defined
at $0.45 \leq z \lesssim 1.3 - 1.9$, the average fractal dimension
is $\langle D \rangle \simeq 1 - 2$. The second region comprises the
scale $1.3 - 1.9 \lesssim z\leq 5.0$ where the fractal dimension
was found to be $\langle D \rangle<1$. These estimates bring initial
confirmation for the theoretical prediction made by Rangel Lemos and
Ribeiro (2008) of an evolving fractal dimension, with decreasing
values for $D$ as $z$ increases. Small values of $D$ mean a more
sparse clustering distribution, which implies that in the past voids
may have dominated the large-scale galactic structure. Our results
also give preliminary indication that $D$ becomes very small, close
to zero, at the outer limits of the FDF survey, a result which
implies that either the galaxies belonging to the fractal system
are not being observed at large values of $z$ or that the
large-scale structure of the universe becomes essentially void
dominated. The latter case perhaps implies that the galactic
clustering itself could have started at a relatively recent epoch in
the evolution of the Universe, when $z<5$. Finally, due to the big
uncertainties in the calculated values of $D$, it is clear these
results must be seen as preliminary, but even so they may also
indicate that the fractal galaxy distribution is possibly better
characterized by more than two scaling ranges in the fractal
dimension, that is, various successive single fractal systems
having several fractal dimensions associated to specific distance
ranges.

The plan of the paper is as follows. In Sect.\ \ref{relfrac} we 
discuss the tools necessary for the fractal analysis of the
galaxy distribution in a relativistic cosmology setting. Sect.\
\ref{fdf} briefly summarizes the features of the FDF survey and
the methodology employed by Ir12a to extract number densities from
the luminosity function built with the FDF galaxies. Sect.\
\ref{results} presents the results of the fractal analysis and
Sect.\ \ref{conclusion} presents our conclusions.

\section{Relativistic fractal cosmology}\lb{relfrac}

\begin{flushright}
\textit{``Theories crumble, but good observations never
fade.''}\\Harlow Shapley
\end{flushright}

The idea that there exists a fractal pattern in the matter distribution
of the Universe is old, actually several centuries old (Ribeiro
2005, Baryshev and Teerikorpi 2002, Gruji\'c 2011, and references
therein). In more recent times, irregularities in the galaxy 
distribution have been studied by several authors since at least the
1900s, that is, well before fractals were introduced in the literature,
as they reasoned that empirical evidence supports the idea
that galaxies clump together to form groups of galaxies, which themselves
form clusters of galaxies and they then form even larger groups, the
galaxy super-clusters, and so on. This was called ``the hierarchical
organization of galaxies'' and models based on this concept were
collectively known as \textit{hierarchical cosmology} (Charlier 1908,
1922; Selety 1922; Einstein 1922; Amoroso Costa 1929; Carpenter 1938;
de Vaucouleurs 1960,
1970; Wertz 1970, 1971; Haggerty and Wertz 1972). Fractal ideas were
developed much later, but soon after their appearance it became clear
that this galactic hierarchical organization amounts to nothing more
than assuming a fractal galaxy distribution (Mandelbrot
1983).\footnote{ There is a clear connection between the late, and most
developed, hierarchical cosmology models (Wertz 1970, 1971) and the
early fractal cosmologies (Pietronero 1987). See Ribeiro (1994) for a
detailed discussion on this topic.} Those earlier studies were,
nevertheless, carried out within the limited scope of the Newtonian
cosmology framework, since relativistic hierarchical (fractal)
cosmologies appeared even later and had first to overcome conceptual
issues like, among others, the meaning of a cosmological fractal
dimension in a curved spacetime whose observations are made
along the past light cone. It is not the aim of this work to
provide a detailed discussion of the conceptual issues surrounding
the relativistic approach to fractal cosmology, discussion which
can be found elsewhere (Ribeiro 1992ab, 1994, 1995, 2001b, 2005;
Rangel Lemos and Ribeiro 2008), although a brief presentation of
these conceptual issues can be found in Sect.\ \ref{theory} below.
So, here we shall mainly restrict ourselves to present the basic
tools capable of providing a fractal description of the galaxy 
distribution in a relativistic cosmology framework.

As discussed above, fractals are characterized by power-laws and,
therefore, we must put forward relativistic-based analytical
tools capable of capturing fractal features from the empirical data,
where the latter are, by definition, collected along the observer's
backward null cone. To elaborate on this point, let us start by
writing the defining expression of the \textit{differential density}
$\gamma$ (Wertz 1970, 1971),
\be
\gamma=\frac{1}{4 \pi {(d)}^2} \frac{\df N}{\df (d)},
\lb{gama1}
\ee 
where $N$ is the cumulative number counts of cosmological sources
(galaxies) and $d$ is the \textit{observational} distance. From this
definition it is clear that $\gamma$ gives the rate of growth in
number counts, or more exactly in their density, as one moves along
the observational distance $d$.

From a relativistic viewpoint, it is well-known that cosmological
distances are not uniquely defined (Ellis 1971, 2007) and, therefore,
we have to replace $d$ for $d_i$ in the equation above, where
the index indicates the chosen distance measure. The ones to be used
here are the \textit{redshift distance} $\dz$, the \textit{luminosity
distance} $\dl$ and the \textit{galaxy area distance} $\dg$. The last
two are connected by the Etherington reciprocity law (Etherington 1933;
Ellis 1971, 2007),
\be
\dl=(1+z)\, \dg,
\lb{eth}
\ee
where $z$ is the redshift. The redshift distance is defined by the
following equation,
\be 
\dz=\frac{c \, z}{H_0},
\lb{red}
\ee
where $c$ is the light speed and $H_0$ is the Hubble constant.
This definition of $\dz$ is, of course, only valid in the FLRW
metric. A07 and Ir12a showed that within the
FLRW cosmology the densities defined with both $\dl$ and $\dz$
have power-law properties and we shall see below that the same
is true with $\dg$. Another distance measure that can be defined
in this context is the \textit{angular diameter distance} $\da$,
also known as area distance. However, densities defined with
$\da$ have the odd behavior of increasing as $z$ increases,
making it unsuitable to use in the context of a fractal analysis
of the galaxy distribution (Ribeiro 2001b, 2005; A07; Rangel
Lemos and Ribeiro 2008).

The discussion above about cosmological distances implies that
Eq.\ (\ref{gama1}) must be rewritten as follows (Ribeiro 2005),
\be
\gamma_i=\frac{1}{4 \pi {(d_i)}^2} \frac{\df N}{\df (d_i)}
        =\frac{\df N}{\df z} {\left[ 4 \pi {(d_i)}^2 \frac{\df
         (d_i)}{\df z} \right]}^{-1},
\lb{gama2}
\ee 
where ($i={\sty G}$, ${\sty L}$, ${\sty Z}$) according to the
distance definition used to calculate the differential density.
Integrating the equation above over an observational volume
$V_i$ produces the \textit{integral density} $\gamma_i^\ast$, which
can be written as (Ribeiro 2005),
\be
\gamma_i^\ast=\frac{1}{V_i} \int_{V_i} \gamma_i \, \df V_i,
\lb{gstar}
\ee
where, 
\be
V_i=\frac{4}{3} \pi {(d_i)}^3.
\lb{volume}
\ee
Clearly $\gamma_i^\ast$ gives the number of sources per unit of
observational volume located inside the observer's past light cone
out to a distance $d_i$. From its definition it is straightforward
to conclude that the following expression holds,
\be
\gamma_i^\ast=\frac{N}{V_i}.
\lb{gstar2}
\ee

One should note that $\gamma$ and $\gamma^\ast$ are \textit{radial}
quantities and, therefore, must not be confused with the similar
looking functions advanced by Pietronero (1987), the conditional
density $\Gamma$ and the integrated conditional density $\Gamma^\ast$
(see also Sylos Labini et al.\ 1998; Gabrielli et al.\ 2005). The
latter two are quantities defined in statistical sense, which means
averaging all points against all points, whereas $\gamma$ and
$\gamma^\ast$ are radial only quantities. Luminosity functions
computed from redshift surveys data are presented as radial
functions, so one should use radial densities with LF derived data.

The \textit{key hypothesis} behind the assumption that the
smoothed-out galaxy distribution forms a single fractal system can
be translated into a simple equation relating the cumulative number
counts of \textit{observed} cosmological sources $\obs{N}$ and the
observational distances $d_i$. This is the \textit{number-distance
relation}, whose expression may be written as, 
\be
\obs{N}=B \, {(d_i)}^D,
\lb{nd}
\ee
where $B$ is a positive constant and $D$ is the fractal dimension.
This expression is the keystone of the \textit{Pietronero-Wertz
hierarchical (fractal) model} (Ribeiro 1994). Note that since
$\obs{N}$ is a cumulative quantity, if, for whatever
reason, beyond a certain distance there are no longer galaxies then
$\obs{N}$ no longer increases with distance. If instead objects are
still detected and counted then it continues to increase. Observational
effects can possibly affect its rate of growth leading to an
intermittent behavior, nevertheless, as $\obs{N}$ is an integral
quantity it must grow or remain constant and thus the exponent in
Eq.\ (\ref{nd}) must be positive or zero.

Substituting the expression above in Eqs.\ (\ref{gama2}) and
(\ref{gstar2}) we easily obtain two forms for the
\textit{de Vaucouleurs density power-law} (Ribeiro 1994), 
\begin{equation}
\obs{\gamma_{i}} = \frac{DB}{4\pi}{(d_{i})}^{D-3},
\lb{gama3}
\end{equation}
\begin{equation}
\obs{\gamma_{i}^{\ast}} = \frac{3B}{4\pi}{(d_{i})}^{D-3}.
\lb{gstar3}
\end{equation}
Thus, if the observed galaxy distribution behaves as a fractal
system with $D<3$, both the observed differential and integral
densities must behave as decaying power-laws. If $D=3$ the
distribution is \textit{observationally} homogeneous, as both
densities become constant and independent of distance.\footnote{ As
extensively discussed by Rangel Lemos and Ribeiro (2008; see also
Ribeiro 2001b, 2005), observational and spatial homogeneity are 
very different concepts in relativistic cosmology. One may have a
cosmological-principle-obeying spatially homogeneous cosmological
model exhibiting observational inhomogeneity, and the other way
round.} The ratio between these two densities yields (Ribeiro 1995),
\begin{equation}\label{directD}
\frac{\obs{\gamma_{i}}}{\obs{\gamma_{i}^{\ast}}}=\frac{D}{3},
\end{equation}
providing a direct method for measuring $D$. If the distribution
is observationally homogeneous then this ratio must be equal to
one. An irregular distribution forming a single fractal system
will have $0 \leq \left( {\obs{\gamma_i}} \big. \big/ {\obs{\gamma_i^\ast}}
\right)<1$. 

\subsection{Observer's past light cone}\lb{theory}

It is very important to stress that the quantities discussed above are
relativistic-based tools defined along the observer's past light cone
null hypersurface. Therefore, even in the FLRW spatially homogeneous
cosmological model these quantities are defined in a different spacetime
manifold foliation than the one where the local density is, by definition,
constant. To see this, one must remember that in relativistic cosmology
all observational quantities are dependent on the coordinates' dynamics.
Thus, the radial number density $n$ depends on both the time and radius
coordinates, that is, $n=n(t,r)$, which reduces to $n(t_0,r)=n_0$ in the
present time surface $t_0$. It is a well-known result that $n_0$ is
constant in the FLRW cosmology. However, besides the present time
surface $t_0$ one may express the number density along the radial light
cone where both $t$ and $r$ are function of this hypersurface's affine
parameter $u$, such that $t=t(u)$, $r=r(u)$, and we can write the light
cone as $t=t[r(u)]$, or simply $t=t(r)$. This means that along the past light
cone the number density is given as $n=n[t(r),r]$. Since $t(r)$ changes,
so does $n[t(r),r]$. Therefore, $n_0$ and $n[t(r),r]$ are defined in
completely different spacetime manifold surfaces.

The reasoning above implies that all other observational quantities
will also be written in terms of the past light cone $t(r)$. Therefore,
the cumulative number counts is $N=N[t(r),r]$ and the observational
distance measures are expressed as $d_i=d_i[t(r),r]$. Under this
viewpoint, fractality means that $N$ will behave as a power-law along
the light cone because the function $t(r)$ does change along this
surface. Fractality is, hence, a past light cone effect, but that only
occurs at $z$ values high enough because at low redshifts the light
cone effects on the observables are negligible, meaning that at low
redshifts $n[t(r),r] \approx n_0$, that is, at low redshifts one can
drop relativity and use the Newtonian approximation. Therefore, to be
able to probe fractality at deep ranges under a relativistic cosmology
perspective we need a survey starting to at least $z \approx 0.2$. As a
consequence, in FLRW cosmology both $\gamma_i$ and $\gamma_i^*$ will
\textit{not} remain constant even if one drops the fractal hypothesis
given by Eq.\ (\ref{nd}). A very detailed discussion of this topic can
be found in the first sections of Ribeiro (2001b). Rangel Lemos \&
Ribeiro (2008) presented in detail the approximation of the minimum
redshift threshold.

The discussion above also implies that simplified illustrations at
low redshift ranges of the fractal approach to the galaxy distribution
are not applicable to the analysis performed in this paper. The light
cone is an entirely relativistic concept, so huge confusion will
certainly arise if one does not acknowledge the difference between
$n_0$ and $n[t(r),r]$. The FLRW model states that $n(t_0,r)=n_0=$
constant, but this cosmology \textit{also} states that $n[t(r),r]$ is
\textit{not} constant (Ribeiro 1992b). The fractal analysis of this
paper is not in the spacetime region where $n$ is constant, defined
by $t=t_0$, but along the past light cone $t=t(r)$ where $n[t(r),r]
\not=$~constant. It is in the latter region, actually the spacetime
surface where astronomy is made, that fractality may be detected. Thus,
fractality may appear only when one correctly manipulates the FLRW
observational quantities along the observer's past light cone at
ranges where the null cone effects start to be relevant. The FDF
survey used in this paper starts at those ranges.

In summary, one cannot neglect relativistic effects in the whole
analysis of this paper as these effects form the very core of the
present analysis.

\section{Number densities of the FDF redshift survey}\lb{fdf}

Ir12a calculated the differential and integral densities of the FDF
galaxy survey by means of a series of steps involving theoretical
and astronomical considerations. These steps included linking the
LF astronomical data and practice with relativistic cosmology number
counts theory according to the model advanced by Ribeiro and Stoeger
(2003; see also A07).

Ir12a started their analysis from the redshift evolving LF parameters
fitted by G04 and G06 to the FDF dataset using a Schechter analytical
profile over the redshifts of 5558 I-band selected galaxies in the
FORS Deep Field dataset. G04 and G06 showed that the selection in the
I-band is projected to miss less than 10\% of the K-band detected
objects, since the AB-magnitudes of the I-band are half a magnitude
deeper than those of the K-band, out to $z=6$, beyond which the Lyman
break does not allow any signal to be detected in the I-band. In
addition, the I-band selection minimizes biases like dust absorption.
All galaxies in those studies were therefore selected in the I-band
and then had their magnitudes for each of the five blue bands (1500
\AA, 2800 \AA, $u'$, $g'$ and $B$) and the three red ones ($r'$, $i'$
and {$z'$}) computed using the best fitting SED given by their authors'
photometric redshift code convolved with the associated filter
function and applying the appropriate K-correction. The photometric
redshifts were determined by G04 and G06 by fitting template spectra
to the measured fluxes on the optical and near infrared images of the
galaxies.

Using the published LF parameters of the FDF survey, Ir12a computed
selection functions by means of its limited bandwidth version for
given LF fitted by a Schechter analytical profile in terms of
absolute magnitudes, as follows (Ribeiro and Stoeger 2003),
\begin{equation}
\lb{sf}
\psi^{\ssty W}(z) = 0.4 \ln 10 \, \int_{-\infty}^{M^{\ssty W}_{\ssty
lim}(z)}
\phi^\ast(z) 10^{0.4[1+\alpha(z)][M^\ast(z) - \Mbar^{\ssty W}]}
%\times \nonumber \\ \times
\exp \{-10^{0.4[M^\ast(z) - \Mbar_{\ssty W}]}\} \df \Mbar^{\ssty W},
\end{equation}
where $\psi$ is the selection function and the index $W$ indicates the
bandwidth filter in which the LF is being integrated. The redshift
evolution expressions of the LF parameters found by G04 and G06 are,
\begin{eqnarray*}
\phi^{\ast}(z) = \phi^{\ast}_{\ssty 0} \, (1+z)^{B^{\ssty W}}, \\
M^{\ast}(z) = M^{\ast}_0 + A^{\ssty W} \ln (1+z), \\
\alpha(z) = \alpha_{\ssty 0},
\end{eqnarray*}
with $A^{\ssty W}$ and $B^{\ssty W}$ being the evolution parameters
fitted for the different $W$ bands and $M^{\ast}_{\ssty 0}$,
$\phi^{\ast}_{\ssty 0}$ and $\alpha_{\ssty 0}$ the local (z $\approx$
0) values of the Schechter parameters as given in G04 and G06.
Inasmuch as all galaxies were detected and selected in the I-band, we
can have,
\begin{equation}
\label{mlim}
M^{\ssty W}_{\ssty lim}(z) = M^{\ssty I}_{\ssty lim}(z) =
I_{\ssty lim} - 5 \log[\dl (z)] - 25 + A^{\ssty I},
\end{equation}
for a luminosity distance $\dl$ given in Mpc. $I_{\ssty lim}$ is
the limiting apparent magnitude of the I-band of the FDF survey,
being equal to 26.8. Its reddening correction is $A^{\ssty I} =
0.035$. The selection functions were, therefore, obtained by
integrating the LF over the {absolute magnitudes} given by G04 and
G06 in the five blue bands and the three red ones, therefore
producing {comoving number densities corresponding to a}
\textit{volume limited galaxy sample} {defined in equally
spaced redshift bins. It is important to notice that the actual
selection of objects was done in G04 and G06, resulting in flux-limited
datasets. Ir12a merely obtained number densities from the corrected
and best fitted LF parameters, which, as discussed in \S \ref{intro}
above, correspond to volume-limited samples. Such number densities
should be as redshift unbiased as the LF parameters used to obtain
them, which ensures unbiased shapes of the density-vs.-distance
relations and their accurate power-law fits, as will be discussed
in \S \ref{results} below.}

{Fig.\ \ref{magplotz} shows the volume-limited samples
corresponding to the redshift limits in each of the considered
redshift bins for the B-band absolute magnitudes of all galaxies in
the FDF survey, together with the absolute magnitude cuts based on the
completeness limit of the I-band. We notice that an absolute magnitude
cut based on the I-band corresponds to volume-limited samples in the
B-band that are safely inside the formal completeness limit for the
B-band, as well as the bounded limit defined by the faintest B-band
absolute magnitude in the FDF survey which corresponds to an apparent
magnitude of approximately 29.8.}

\begin{figure}[tb]
\centering
\includegraphics[width=12cm]{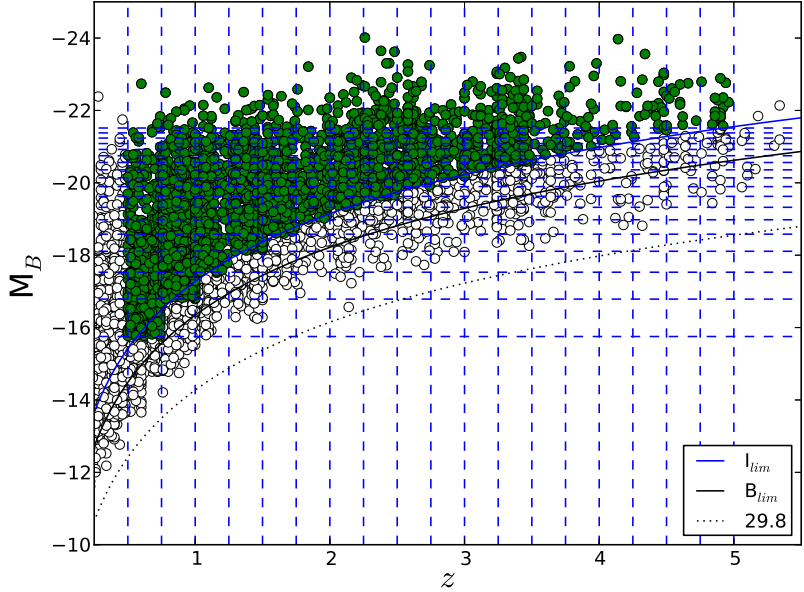}
\caption{Plot of the B-band absolute magnitudes of the FDF galaxy
survey in the redshift range $0.45 \leq z \leq 5$. The green filled
circles represent the FDF flux-limited I-band selected dataset that
would be included in a volume-limited subsample given by the FDF
redshift bins and the I-band absolute magnitude cuts of Ir12a,
represented here as horizontal dashed blue lines.\label{magplotz}}
\end{figure}

The blue bands of G04, in the range $0.5 \leq z \leq 5$,
were combined in two sets, the blue optical bands $g'$ and $B$ and
the blue UV bands 1500 \r{A}, 2800 \r{A} and $u'$. The red-band
dataset of G06, in the range $0.45 \leq z \leq 3.75$, was also
combined in a single set.

The next step consisted in obtaining observational differential
number counts $\obs{\df N/\df z}$ by means of the following
expression discussed in detail in Ir12a,
\begin{equation}
\bigobs{\frac{\df N}{\df z}} = \frac{V_{\ssty \rm C}}{V_{\ssty
       \rm Pr}} \frac{\psi}{n} \frac{\df N}{\df z},
\lb{dNdz}
\end{equation}
where $V_{\ssty \rm C}$ and $V_{\ssty \rm Pr}$ are, respectively,
the comoving and proper volumes, $\df N/\df z$ is the theoretical
differential number counts and $n$ is the number density of radiating
sources in proper volume.  All theoretical quantities, that is,
$V_{\ssty \rm C}$, $V_{\ssty \rm Pr}$, $n$, $\df N/\df z$, were
independently computed in the
FLRW spacetime with $\Omega_{m_{\ssty 0}} = 0.3$,
$\Omega_{\Lambda_{\ssty 0}} = 0.7$, $H_0=70 \; \mbox{km} \;
{\mbox{s}}^{-1} \; {\mbox{Mpc}}^{-1}$ and included in the equation
above, together with the results of $\psi$ previously obtained, to
solve this expression.

Eq.\ (\ref{dNdz}) performs essentially a removal of the
cosmological model assumed by the observers when they calculated the
LF. The aim of this equation is to recover the observed differential
number count $\obs{\df N/\df z}$ used by those who built the LF, since to
do so they had to assume a cosmology. But, this cosmology extraction
does not remove the data corrections, because these were made when
the LF was fitted to the data. Therefore, the final $\obs{\df N/\df z}$
data are not really the raw, observed, data, but the fitted raw
data. The selection function $\psi$ is the observational part coming
directly from the LF, or more specifically, from its integration over
absolute magnitudes. $V_{\ssty \rm C}$ and $V_{\ssty \rm Pr}$ are
just volume transformations since it is nowadays standard practice to
calculate the LF using comoving volume, which has to be removed if we
want to obtain number densities using different volume definitions.
The theoretical differential number count $\df N/\df z$ and theoretical
density $n$ account for the assumed cosmology when the LF was built,
but since $n$ is defined in terms of the proper volume, this fact
also has to be considered in the extraction of the assumed cosmological
model as made by Eq.\ (\ref{dNdz}). This cosmology extraction
procedure is explained in detail in Ir12a.

The additional necessary steps included computing the cosmological
distances $d_i$ in the FLRW model with the same cosmological
parameters above, integrating $\obs{\df N/\df z}$ to obtain
$\obs{N(z)}$ and changing $\obs{\df N/\df z}$ into $\obs{\df
N/\df (d_i)}$. All these results finally allowed the calculation
of both $\obs{\gamma_i}$ and $\obs{\gamma_i^\ast}$ according to
Eqs.\ (\ref{gama2}) and (\ref{gstar2}) in the three observational
sets above.

It is necessary to point out that although this methodology is
capable of extracting from the LF the cosmological model implicitly
assumed in its calculation so that the final observational number
counts becomes model independent, the standard cosmology enters back
into our problem because both $\gamma_i$ and $\gamma_i^\ast$ are
function of the cosmological distances $d_i$, which themselves require
a cosmological model for their evaluation. 

As final remarks, we must emphasize again that this work is
not about inhomogeneity in the comoving number density, but
inhomogeneity defined along the observer's past light cone. Thus,
one can assume a spatial uniform distribution stemming from the
standard cosmological model, which is at the heart of the
$1/V_{\mathrm{max}}$ LF estimator used in the computation of the LF,
obtaining a meaningful $\phi^*(z)$, and a relativistic distribution
along the past light cone which is not uniform as a result of both
expansion effects and the luminosity and/or number density evolution
with the redshift in the LF. Such a difference in the manifold foliation
where our densities are defined is essential in order to understand our
approach. Therefore, \textit{relativistic corrections cannot be ignored
in any part of our analysis and results}, nor can galaxy evolution,
especially at large redshifts as is our case here. Our relativistic
number densities are a convolution between the geometrical effect of
expansion and source evolution, both in the luminosity and number
density evolution probed by the LF. Our aim is to find out if this
convolution produces observed fractality along the past light cone.

\section{Fractal analysis of the FDF survey}\lb{results}

The steps described in the previous section provided data on
$\obs{\gamma_i}$ and $\obs{\gamma_i^\ast}$ in the three combined
observational bands, blue optical, blue UV and red. Once in possession
of these results, as well as the ones for $d_i \, (z)$ in a FLRW 
cosmology, we were able to carry out a fractal analysis of the FDF
galaxy distribution data by testing their fractal compatibility
according to the expressions described in Sect.{~}\ref{relfrac}.

\subsection{Direct calculation of the fractal dimension}\lb{direct}

The simplest test is to calculate the fractal dimension by means
of Eq.\ (\ref{directD}). Figure \ref{rates} shows graphs of $D$
versus the redshift, where the fractal dimension is estimated by
the ratio $D=3\obs{\gamma_i} \, / \, \obs{\gamma_i^\ast}$. The
error bars, obtained by standard quadratic propagation, are big.
Even so, some conclusions can be drawn from the plots.

Firstly, as predicted by Rangel Lemos and Ribeiro (2008), the
fractal dimension decreases as the redshift increases which suggests
the absence of an unique fractal dimension at the sample's redshift
intervals. In other words, an unique single fractal system does not
seem to be a good approximation to describe the FDF galaxy distribution,
since, if that were the case, according to Eq.\ (\ref{directD}) the
graphs in Fig.\ \ref{rates} would have to show an approximate
horizontal line indicating a constant fractal dimension. However,
due to the big uncertainties such a situation cannot yet be entirely
ruled out, although an unique single fractal description seems unlikely. 

Secondly, the homogeneous case $D=3$ occurs only very marginally, at
the top of very few error bars. Except for a single plot, the red
galaxies calculated using $\dg$, all others graphs suggest $D
\lesssim 2$ in most of the studied redshift interval. There are very
few instances where the top of some error bars show $D>3$, but
a fractal system embedded in a three-dimensional topological space
cannot have its fractal dimension bigger than the topological
dimension and, hence, such values ought to be dismissed. Similarly,
the bottom of some error bars reach $D<0$, but as we have discussed
above such results are not valid because the number counts is an
integral quantity and its exponent in Eq.\ (\ref{nd}) is either
positive or zero and, therefore, these results ought to be
dismissed as well.  Thus, considering the error bars the fractal
dimension is bounded to its maximum allowed range, $0 \leq D \leq 3$,
but the plots indicate an apparent asymptotic tendency
towards $D=0$.

\subsection{Calculation of $D$ by power-law fitting}\lb{power-law}

Eqs.\ (\ref{gama3}) and (\ref{gstar3}) show that both densities
should follow a power-law pattern if the galaxy distribution can
really be described as a fractal system. Then, performing linear
fits in the logarithmic plots of $\obs{\gamma_i}$ and
$\obs{\gamma_i^\ast}$ against $d_i$ will provide values for $D$.
The simplest approach for a fractal description of the galaxy
distribution after we dismiss the single fractal approximation
is a system with two scaling ranges in the fractal dimension,
that is, two consecutive single fractal systems with different
fractal dimensions at successive distance ranges.

Next we show the results of a two-straight-lines fit to the data.

\subsubsection{Differential density $\obs{\gamma_i}$}\lb{gama}

Fig.\ \ref{gamas} shows the plots of all differential densities
defined in the three cosmological distances used here against
their respective distances. Clearly it is possible to fit two
straight lines to the data, whose slopes at different redshift
intervals provide values for $D$ by means of Eq.\ (\ref{gama3}).
For best fit results, the redshift range can be divided in two
intervals, the first being $0.45 \leq z \lesssim 1.3-1.9$ and
the second one in the range $1.3-1.9 \lesssim z \leq 5.0$. Let
us call the former as \textit{region~I} and the latter as
\textit{region II}.

The values of D calculated in region I by means of $\obs{\gamg}$,
$\obs{\gaml}$ and $\obs{\gamz}$ basically agree with one another
in their respective redshift intervals and within the error
margins. However, all fractal dimension values obtained in region
II are negative and mostly outside the bounds established by the
direct method discussed in Sect.\ \ref{direct} above, whereas the
results in region I are within those bounds. Negative fractal
dimensions ought to be dismissed since they are not defined in
the context discussed here (see the discussion after Eq.\
(\ref{nd}) above) and, therefore, only the valid results are
summarized in Tables \ref{table1} and \ref{table1a}.

The spurious values of the fractal dimension in region II comes
from the fact that, by definition, the differential densities
measure the rate of growth in number counts, as $\gamma_i
\propto \df N/ \df z$ (see Eq.\ \ref{gama2}). Inasmuch as 
$\df N/ \df z$ increases, reaches a maximum and then decreases,
this behavior substantially enhances the decline in $\gamma$
when $\df N/ \df z$ is evaluated at redshift values beyond
its maximum. In addition, by measuring a rate of growth in
number counts, $\gamma$ is much more sensitive to local
fluctuations and noisy data. Thus, the steep decline
detected in the slopes of the fitted lines in region II of
the $\obs{\gamma_i} \times d_i$ plots are a consequence of
these distortion effects at the redshift limits of the
sample, resulting then in spurious negative values for $D$.

The reasoning above being true, we should then expect the
absence of such bogus negative fractal dimension values
when they are calculated with the integral densities in similar
$\obs{\gamma_i^\ast} \times d_i$ plots, because $\gamma_i^\ast
\propto N$ (see Eq.\ \ref{gstar2}). As the cumulative number
counts $N$ only grows or stays constant, describing therefore
the change in number counts for the entire observational volume,
this property also renders $\gamma_i^\ast$ less sensitive to
tail fluctuations. Hence, $\gamma_i^\ast$ should not present
an enhanced decline distortion at the tail of the
distribution and the values for $D$ obtained with $\gamma_i^\ast$
should also not assume phony negative values. As we shall see
below this is what really happens.

\subsubsection{Integral density $\obs{\gamma_i^\ast}$}\lb{gstarsec}

The same division in two regions was assumed in order to fit
straight lines to the data plots of the integral density versus
their respective cosmological distances. Fig.\ \ref{gamasstar} shows
the $\obs{\gamma_i^\ast} \times d_i$ plots where the fractal dimension
was calculated by estimating the power-law exponent as given in
Eq.\ (\ref{gstar3}). The results are summarized in Tables 
\ref{table2} and \ref{table2a}.

The calculated figures show an absence of negative values for the
fractal dimension in region II, even considering the error margins,
as predicted above.  Besides, all results are well within the
bounds established in Sect.\ \ref{direct}. Thirdly, although the
values of $D$ obtained from the $\obs{\gamma_i^\ast} \times d_i$
plots in region I are somewhat higher than those obtained in the
same region by the $\obs{\gamma_i} \times d_i$ plots, they are
consistent, or very closely consistent, with each other
considering the calculated uncertainties. This reinforces the
view that the results for $D$ obtained in region II from the
$\obs{\gamma_i} \times d_i$ plots are indeed spurious, especially
nearby the limits of the sample. 

\subsection{Discussion}\lb{dis}

In order to better examine the results above, let us calculate
averages for the fractal dimensions in regions I and II for all
galaxy types but, specifying if they were obtained by the
differential or integral number densities. These averages are as
follows,
\be
  {\langle D \rangle}_I^\gamma=0.8^{\ssty+0.7}_{\ssty -0.7}
   \: , \; \; 
  {\langle D \rangle}_I^{\gamma^\ast}=1.4^{\ssty +0.7}_{\ssty -0.6}
   \: , \; \; 
  {\langle D \rangle}_{II}^{\gamma^\ast}=0.5^{\ssty +1.2}_{\ssty -0.4}
   \: .  
\label{avg}
\ee
We have dismissed the result for ${\langle D \rangle}_{II}^\gamma$
due to its spurious nature, as discussed above. We note that due to
the data diversity and limitation, that is, different types of
galaxies and an analysis of a single survey which probed a very
limited part of the sky, these results should be considered only
as general estimates, but they allow us to reach some conclusions.

Firstly, it is clear that we can consider the galaxy distribution
as being described by a bi-fractal system,\footnote{ A fractal
system with two scaling ranges in the fractal dimension is called
as `bi-fractal.' However, this term
%may lead to ambiguous interpretation because it
is also sometimes used to name a fractal
system that simultaneously has two fractal dimensions in the same
scaling range, that is, a system of multifractal nature. In this
paper we use the term `bi-fractal' to convey the
first definition above.} at least as far as the FDF data is
concerned. Secondly, despite being different, the values of
${\langle D \rangle}_{I}^\gamma$ and ${\langle D
\rangle}_{I}^{\gamma^\ast}$ agree with one another within the
error margins. This allows us to reach a third conclusion, which
is that up to $z \sim 1.5$ the fractal dimension is probably in
the range $D=1-2$, whereas for $ 1.5 \lesssim z \lesssim 5.0$ we
probably have $D=0-1$. It is also clear that the integral density
provides a much better tool for estimating the fractal dimension,
since it does not produce bogus negative values for $D$ at higher
redshifts. Finally, the results show that a fractal analysis of
the large-scale galaxy distribution could potentially bring
insights in its evolution as $D$ could provide a parameter for
void evolution. This is so because a decreasing fractal dimension
at increasing redshift ranges indicates that in the past galaxies
and galaxy clusters were much more sparsely distributed than at
recent epochs, possibly meaning a more dominant role for voids
in the large-scale galactic structure at those earlier times.

\section{Conclusions}\lb{conclusion}

In this paper we have performed a fractal analysis of the galaxy
distribution of the FORS Deep Field (FDF) galaxy redshift survey
in the range $0.45 \leq z \leq 5.0$ under the assumption that
this distribution forms a fractal system. The cosmological distances
$d_i$ and their respective observed differential and integral number
densities $\obs{\gamma_i}$ and $\obs{\gamma_i^\ast}$ were used to
calculate the fractal dimension $D$ of the fractal galactic system
by two methods: the direct calculation, through the expression $D=3
\obs{\gamma_i} / \obs{\gamma_i^\ast}$, and by linear fitting, to
extract $D$ from the exponents of the power-laws formed by the plots
$\obs{\gamma_i} \times d_i$ and $\obs{\gamma_i^\ast} \times d_i$. The
index $i$ stands for ($i={\sty G}$, ${\sty L}$, ${\sty Z}$) according
to the three cosmological distances used in this paper, the galaxy
area distance $\dg$, the luminosity distance $\dl$ and the redshift
distance $\dz$. We have used the observed number densities
$\obs{\gamma_i}$ and $\obs{\gamma_i^\ast}$ previously calculated by
Iribarrem et al.\ (2012a) in the standard FLRW cosmological model
with $\Omega_{m_{\ssty 0}} = 0.3$, $\Omega_{\Lambda_{\ssty 0}} =
0.7$ and $H_0=70 \; \mbox{km} \; {\mbox{s}}^{-1} \; {\mbox{Mpc}}^{-1}$
using the luminosity function parameters of the FDF survey as
computed by Gabasch et al.\ (2004, 2006) by means of a Schechter
analytical profile. Both $\obs{\gamma_i}$ and $\obs{\gamma_i^\ast}$
were computed in the three sets of combined galaxy types adopted
by Ir12a, namely blue optical, blue UV and red galaxies, and a cut
in absolute magnitudes was used to select the galaxies that entered
in the computation of both quantities.

Although the adopted galaxy sample probed a limited part of the
sky, it has the advantage of being deep enough for the inhomogeneous
irregularities of the galaxy distribution to be detected along
the past light cone even in the spatially homogeneous standard
FLRW cosmological model adopted here. These inhomogeneities are
better detected by $\obs{\gamma^\ast}$, since $\obs{\gamma}$ is
subject to an important distortion leading to a steep decline in
its computed values at high redshift values, an effect which
renders the results obtained with $\obs{\gamma}$ more error prone.

The direct calculation of $D$ produced results within the allowed
boundaries of the fractal dimension, $0 \leq D \leq 3$, when error bars are
considered, but suggested an asymptotic tendency towards $D=0$
as $z$ increases. This direct method also showed \textit{(i)} an
evolution of the fractal dimension, since $D$ decreases as $z$
increases, \textit{(ii)} that the homogeneous case $D=3$ is only
marginally obtained even at low redshift values and \textit{(iii)}
that an unique single fractal system encompassing the whole
redshift range of the FDF sample is not a good approximation to
describe the FDF galaxy distribution. 

Calculating the fractal dimension by means of the exponent of the
power-laws formed by the $\obs{\gamma_i} \times d_i$ and
$\obs{\gamma_i^\ast} \times d_i$ plots showed that the best fits
were obtained by considering the galaxy distribution as being
bi-fractal, that is, characterized by two scaling ranges in the
fractal dimension. In other words, by bi-fractal we mean two
fractal regimes, or two single fractal systems, at different
and successive ranges. The first set of values for the fractal
dimension were calculated in the range $0.45 \leq z \lesssim
1.3-1.9$, named as region I, whereas the second set of values
for $D$, named region II, was defined by the redshift range
$1.3-1.9 \lesssim z \leq 5.0$. Average results indicated that
the fractal dimension varies from $D=2$ to $D=1$ in region I
and from $D=1$ to $D=0$ in region II. Such evolution of the
fractal dimension could provide insights on how the large-scale
galactic structure evolves, since these results suggest that in
the past individual galaxies and galactic clusters were much
more sparsely distributed than at later epochs and, therefore,
the Universe was then possibly dominated by voids.

Finally, it is worth mentioning that Iribarrem \etal (2013)
fitted the luminosity function in a Lema\^{i}tre-Tolman-Bondi
(LTB) spatially inhomogeneous cosmological model using over 10,000
sources of the Herschel/PACS evolutionary probe (PEP) survey in
the observer's far-infrared passbands of 100 $\mu$m and 160
$\mu$m. Then Iribarrem \etal (2014) used those results to obtain
power-law fits for both $\obs{\gaml}$ and $\obs{\gaml^\ast}$ in
the high redshift range $1.5 \lesssim z \lesssim 3.2$ (see their
Fig.\ 6). Although Iribarrem \etal (2014) focused on other issues,
one can infer from their results for $\obs{\gaml^\ast} \times \dl$ in both
100 $\mu$m and 160 $\mu$m passbands that they produced $\langle D \rangle
=0.6 \pm 0.1$, that is, an average value comparable to the high
redshift fractal dimension found in the red region of the FDF
survey studied in this paper using the FLRW cosmology (see table
\ref{table2} below).

\vspace{0.5cm}

{\small We thank A.\ R.\ Lopes for kindly providing her results
obtained with the FDF data and for useful discussions. We also
thank four referees for their comments and useful suggestions
which improved the paper. G.\ C.-S.\ and A.\ I.\ are grateful
to the Brazilian agency CAPES for financial support.}

\begin{table}
\caption{Fractal dimensions calculated using $\obs{\gaml}$ and
$\obs{\gamz}$} 
\centering
\begin{tabular}{|c|lll}
\hline
galaxies &
\multicolumn{1}{c|}{$z$} &
\multicolumn{1}{c|}{$\obs{\gaml} \times \dl$} &
\multicolumn{1}{c|}{$\obs{\gamz} \times \dz$} \\
\hline
blue optical &
\multicolumn{1}{c|}{$0.5-1.2$} &
\multicolumn{1}{l|}{$D=0.6\pm0.3$} &
\multicolumn{1}{l|}{$D=0.7\pm0.4$} \\ &
\multicolumn{1}{c|}{$1.3-5.0$} &
\multicolumn{1}{c|}{$-$} &
\multicolumn{1}{c|}{$-$} \\ &
\multicolumn{1}{c|}{} &
\multicolumn{1}{c|}{} &
\multicolumn{1}{c|}{}  \\
blue UV &
\multicolumn{1}{c|}{$0.5-1.2$} &
\multicolumn{1}{l|}{$D=0.4\pm0.3$} &
\multicolumn{1}{l|}{$D=0.6\pm0.3$} \\ &
\multicolumn{1}{c|}{$1.3-5.0$} &
\multicolumn{1}{c|}{$-$} &
\multicolumn{1}{c|}{$-$} \\ &
\multicolumn{1}{c|}{} &
\multicolumn{1}{c|}{} &
\multicolumn{1}{c|}{} \\
red  &
\multicolumn{1}{c|}{$0.45-1.15$} &
\multicolumn{1}{l|}{$D=0.8\pm0.3$} &
\multicolumn{1}{l|}{$D=1.0\pm0.3$} \\ &
\multicolumn{1}{c|}{$1.25-3.75$} &
\multicolumn{1}{c|}{$-$} &
\multicolumn{1}{c|}{$-$} \\
\hline
\end{tabular}
\lb{table1}
\end{table}

\begin{table}
\caption{Fractal dimensions calculated using $\obs{\gamg}$} 
\centering
\begin{tabular}{|c|ll}
\hline
galaxies &
\multicolumn{1}{c|}{$z$} &
\multicolumn{1}{c|}{$\obs{\gamg} \times \dg$} \\
\hline
blue optical &
\multicolumn{1}{c|}{$0.5-1.8$} &
\multicolumn{1}{l|}{$D=1.0\pm0.3$} \\ &
\multicolumn{1}{c|}{$1.9-5.0$} &
\multicolumn{1}{c|}{$-$} \\ &
\multicolumn{1}{c|}{} &
\multicolumn{1}{c|}{} \\
blue UV &
\multicolumn{1}{c|}{$0.5-1.8$} &
\multicolumn{1}{l|}{$D=0.8\pm0.3$} \\ &
\multicolumn{1}{c|}{$1.9-5.0$} &
\multicolumn{1}{c|}{$-$} \\ &
\multicolumn{1}{c|}{} &
\multicolumn{1}{c|}{} \\
red  &
\multicolumn{1}{c|}{$0.45-1.75$} &
\multicolumn{1}{l|}{$D=1.3\pm0.3$} \\ &
\multicolumn{1}{c|}{$1.85-3.75$} &
\multicolumn{1}{c|}{$-$} \\
\hline
\end{tabular}
\lb{table1a}
\end{table}

\begin{table}
\caption{Fractal dimensions calculated using $\obs{\gaml^\ast}$ and
$\obs{\gamz^\ast}$} 
\centering
\begin{tabular}{|c|lll}
\hline
galaxies &
\multicolumn{1}{c|}{$z$} &
\multicolumn{1}{c|}{$\obs{\gaml^\ast} \times \dl$} &
\multicolumn{1}{c|}{$\obs{\gamz^\ast} \times \dz$} \\
\hline
blue optical &
\multicolumn{1}{c|}{$0.5-1.2$} &
\multicolumn{1}{l|}{$D=1.1\pm0.3$} &
\multicolumn{1}{l|}{$D=1.6\pm0.4$} \\ &
\multicolumn{1}{c|}{$1.3-5.0$} &
\multicolumn{1}{l|}{$D=0.3\pm0.1$} &
\multicolumn{1}{l|}{$D=0.4\pm0.1$} \\ &
\multicolumn{1}{c|}{} &
\multicolumn{1}{c|}{} &
\multicolumn{1}{c|}{}  \\
blue UV &
\multicolumn{1}{c|}{$0.5-1.2$} &
\multicolumn{1}{l|}{$D=1.1\pm0.3$} &
\multicolumn{1}{l|}{$D=1.3\pm0.3$} \\ &
\multicolumn{1}{c|}{$1.3-5.0$} &
\multicolumn{1}{l|}{$D=0.3\pm0.1$} &
\multicolumn{1}{l|}{$D=0.3\pm0.1$} \\ &
\multicolumn{1}{c|}{} &
\multicolumn{1}{c|}{} &
\multicolumn{1}{c|}{} \\
red  &
\multicolumn{1}{c|}{$0.45-1.15$} &
\multicolumn{1}{l|}{$D=1.2\pm0.3$} &
\multicolumn{1}{l|}{$D=1.5\pm0.4$} \\ &
\multicolumn{1}{c|}{$1.25-3.75$} &
\multicolumn{1}{l|}{$D=0.5\pm0.2$} &
\multicolumn{1}{l|}{$D=0.6\pm0.2$} \\
\hline
\end{tabular}
\lb{table2}
\end{table}

\begin{table}
\caption{Fractal dimensions calculated using $\obs{\gamg^\ast}$} 
\centering
\begin{tabular}{|c|ll}
\hline
galaxies &
\multicolumn{1}{c|}{$z$} &
\multicolumn{1}{c|}{$\obs{\gamg^\ast} \times \dg$} \\
\hline
blue optical &
\multicolumn{1}{c|}{$0.5-1.8$} &
\multicolumn{1}{l|}{$D=1.6\pm0.3$} \\ &
\multicolumn{1}{c|}{$1.9-5.0$} &
\multicolumn{1}{l|}{$D=0.7\pm0.5$} \\ &
\multicolumn{1}{c|}{} &
\multicolumn{1}{c|}{} \\
blue UV &
\multicolumn{1}{c|}{$0.5-1.8$} &
\multicolumn{1}{l|}{$D=1.5\pm0.2$} \\ &
\multicolumn{1}{c|}{$1.9-5.0$} &
\multicolumn{1}{l|}{$D=0.5\pm0.4$} \\ &
\multicolumn{1}{c|}{} &
\multicolumn{1}{c|}{} \\
red  &
\multicolumn{1}{c|}{$0.45-1.75$} &
\multicolumn{1}{l|}{$D=1.8\pm0.3$} \\ &
\multicolumn{1}{c|}{$1.85-3.75$} &
\multicolumn{1}{l|}{$D=1.0\pm0.7$} \\
\hline
\end{tabular}
\lb{table2a}
\end{table}

%\clearpage

\begin{figure}
\begin{center}$
\begin{array}{ccc}
\includegraphics[width=5.3cm]{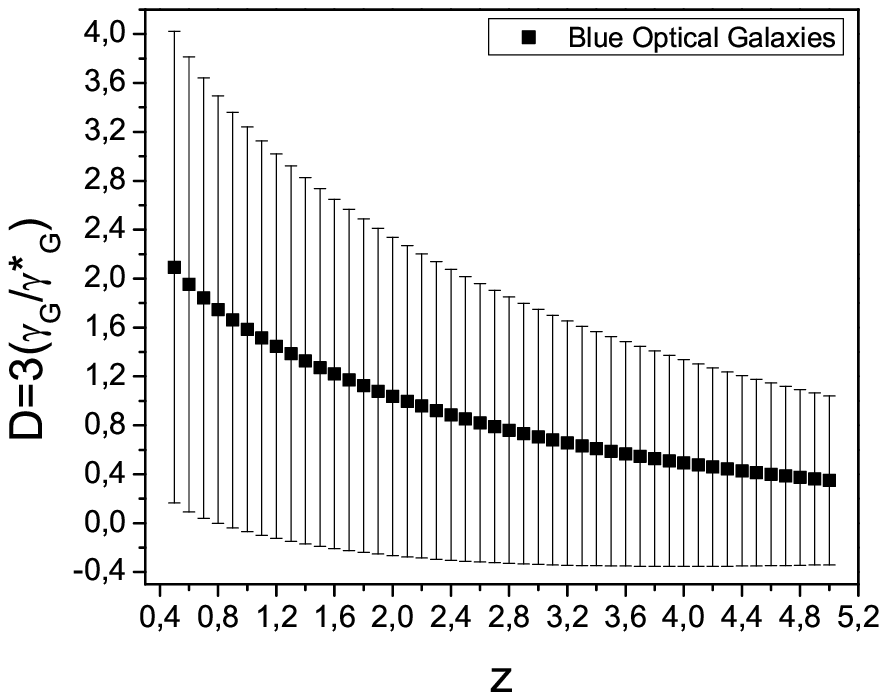} &
\includegraphics[width=5.3cm]{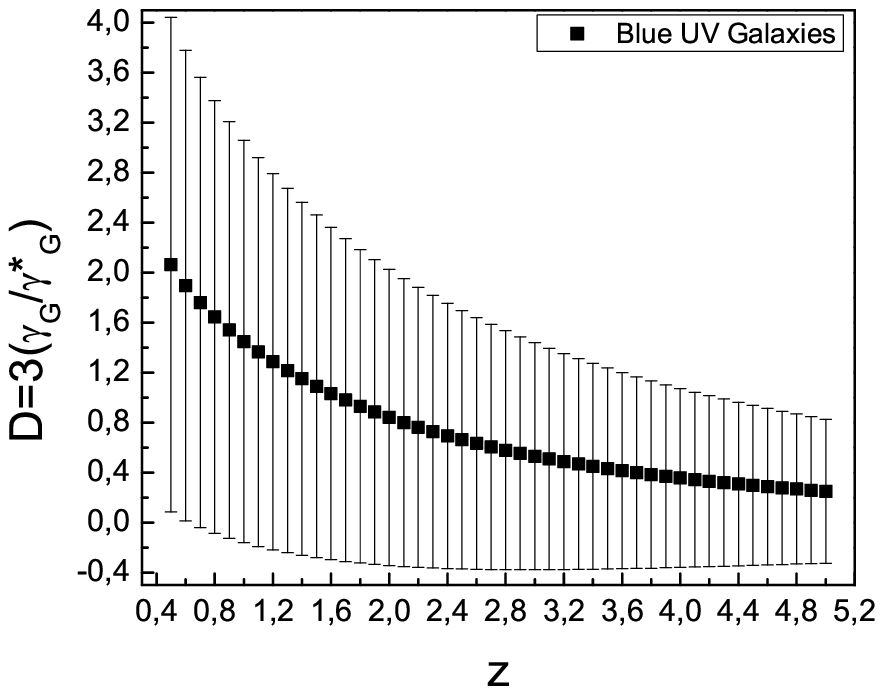} &
\includegraphics[width=5.3cm]{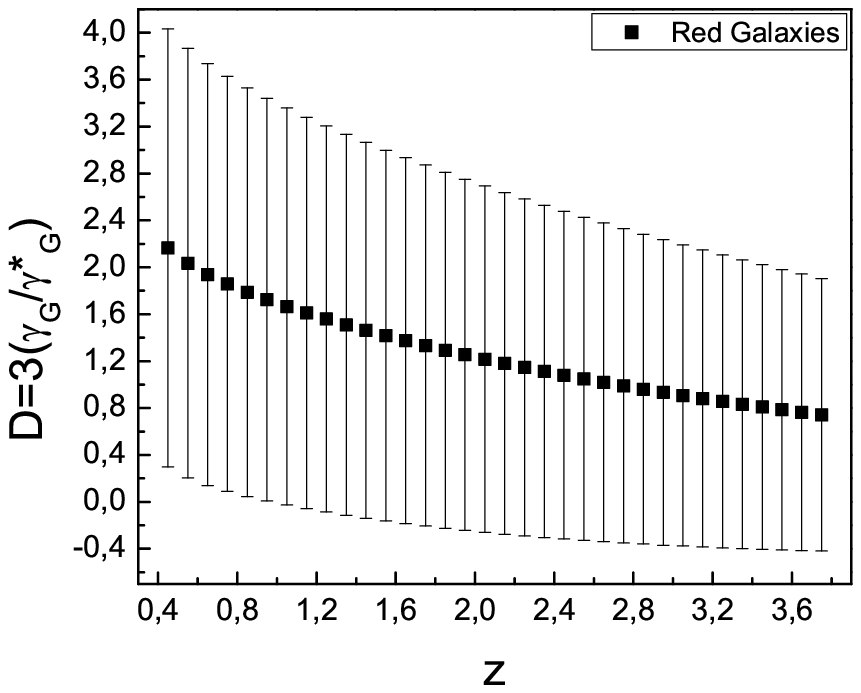} \\
\includegraphics[width=5.3cm]{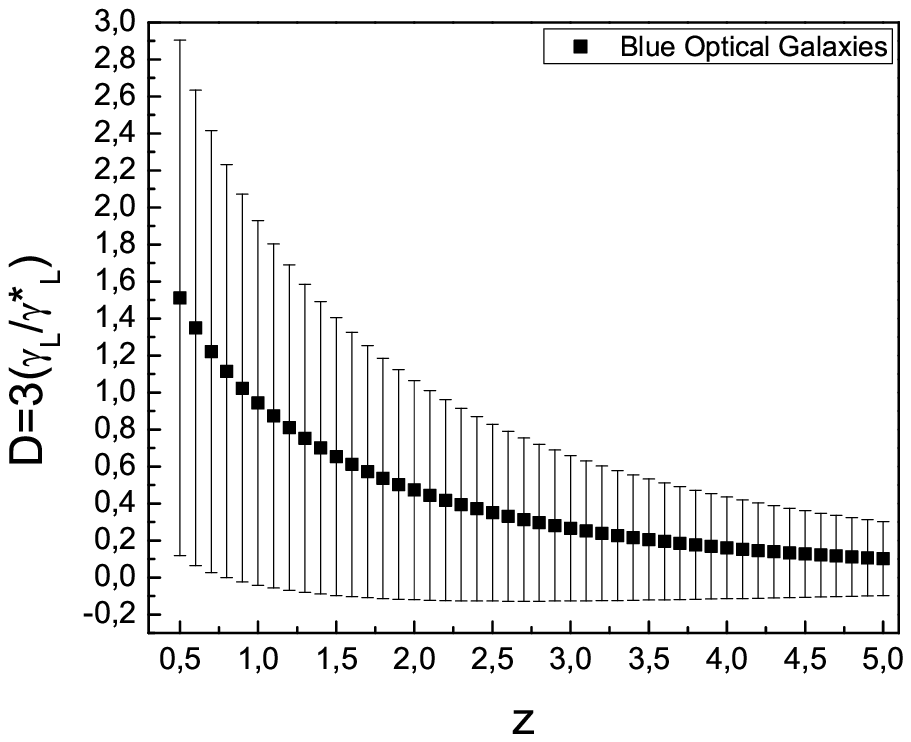} &
\includegraphics[width=5.3cm]{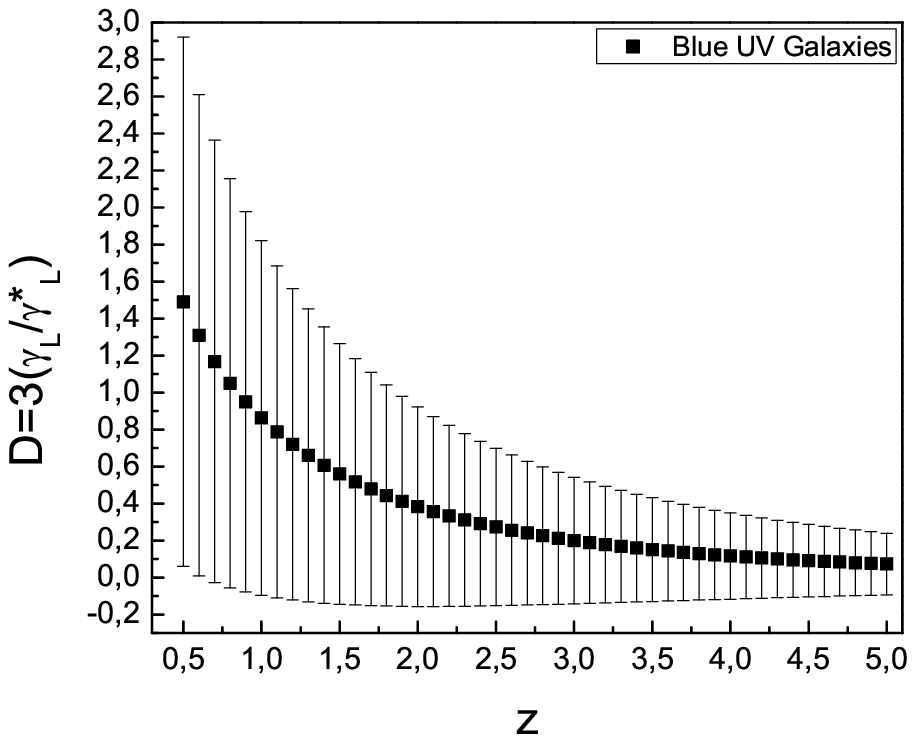} &
\includegraphics[width=5.3cm]{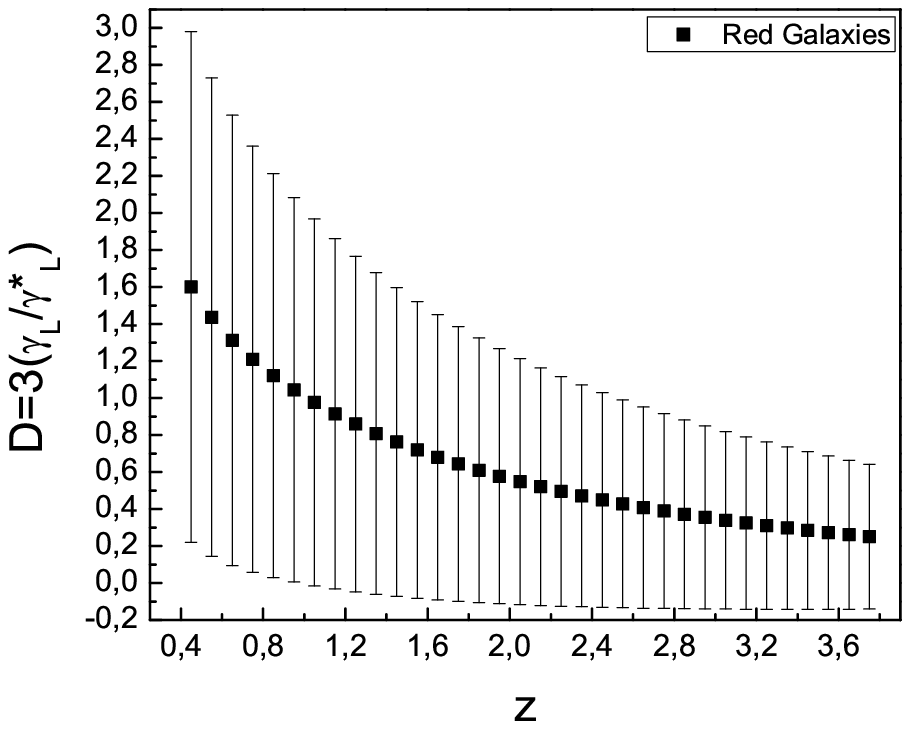} \\
\includegraphics[width=5.3cm]{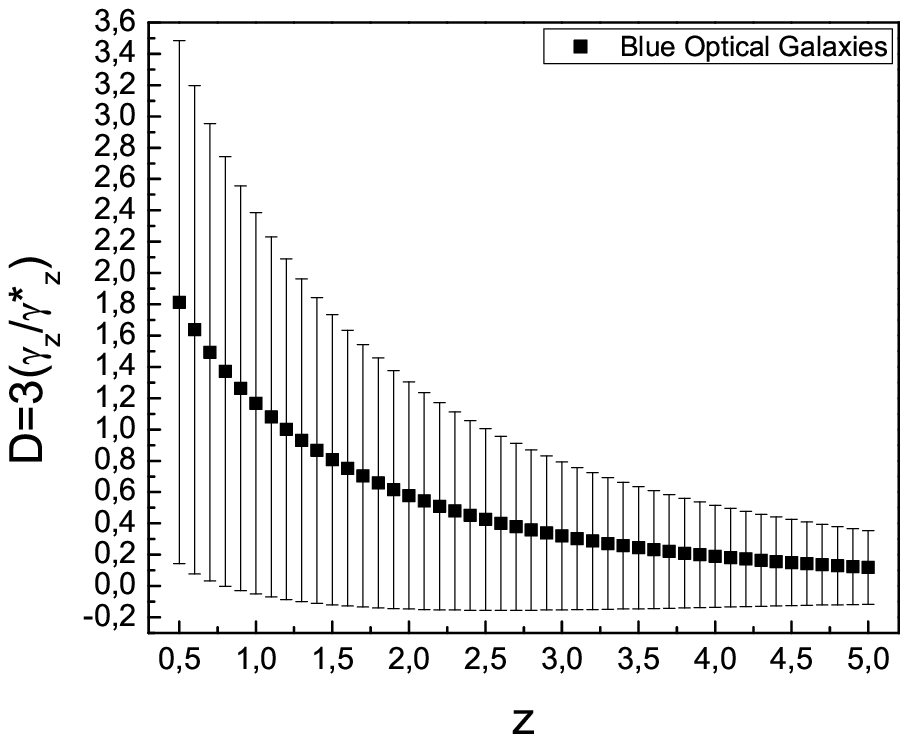} &
\includegraphics[width=5.3cm]{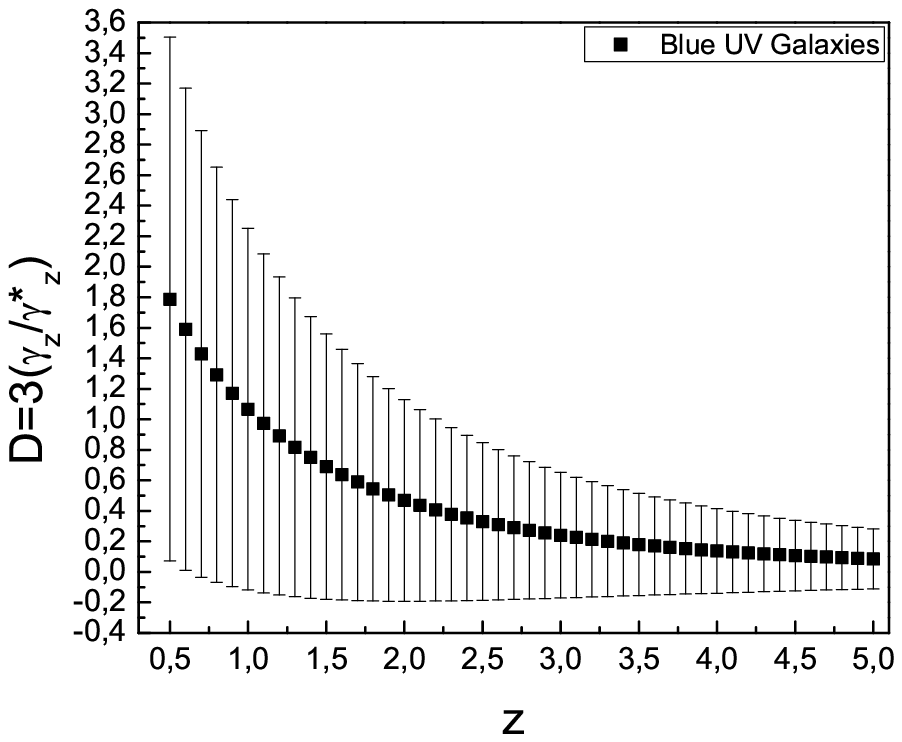} &
\includegraphics[width=5.3cm]{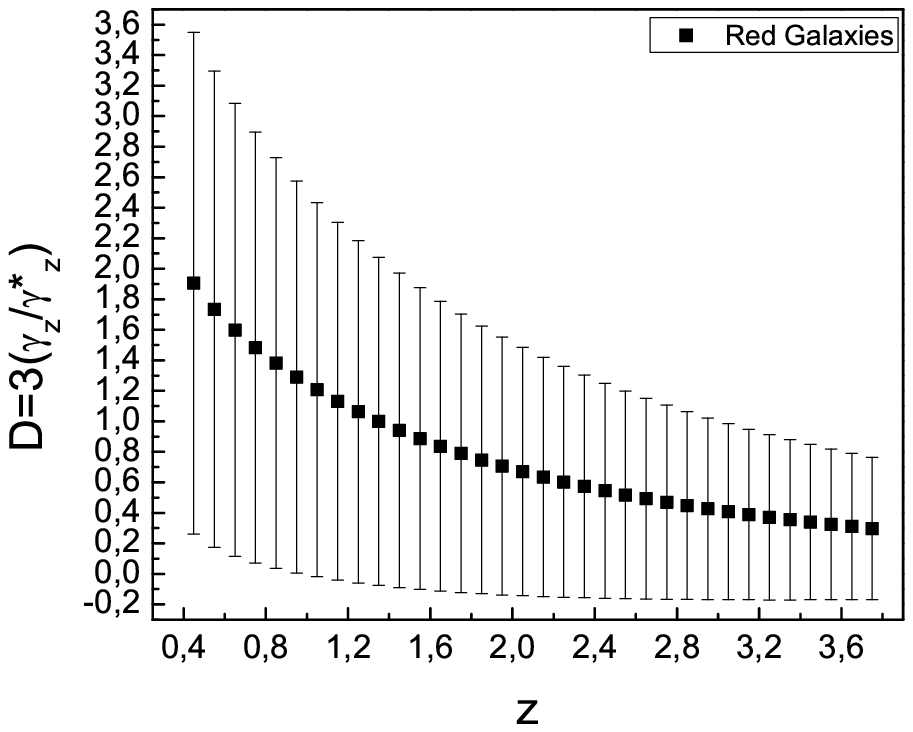} 
\end{array}$
\end{center}
\caption{Fractal dimensions calculated with $\obs{\gamma_i}$
and $\obs{\gamma_i^\ast}$ using Eq.\ (\ref{directD}).}
\lb{rates}
\end{figure}

\begin{figure}
\begin{center}$
\begin{array}{ccc}
\includegraphics[width=5.3cm]{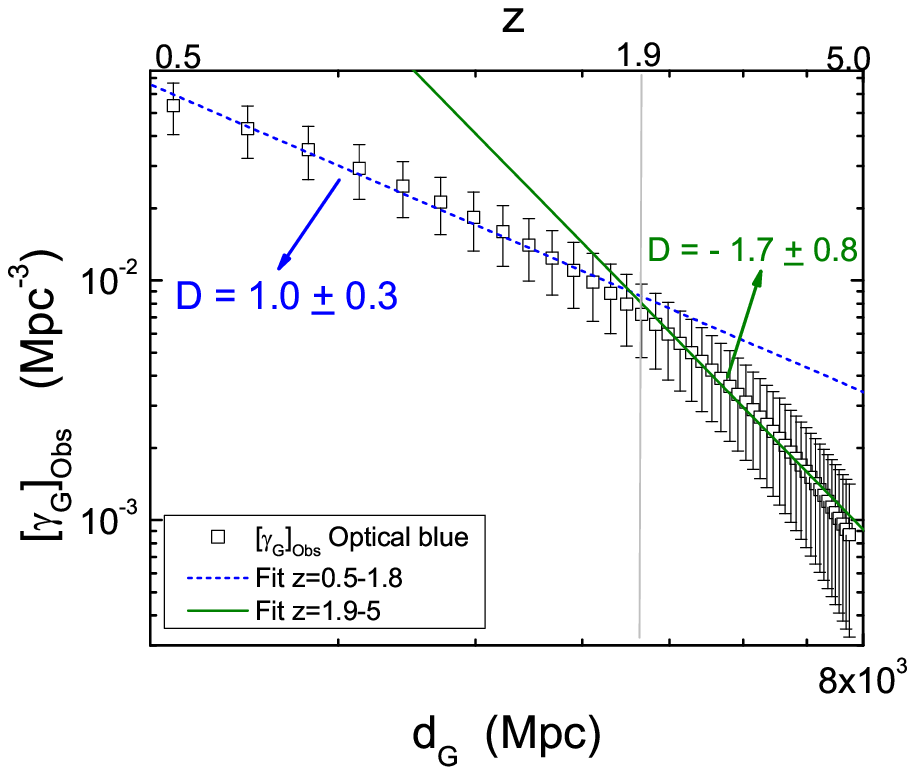} &
\includegraphics[width=5.3cm]{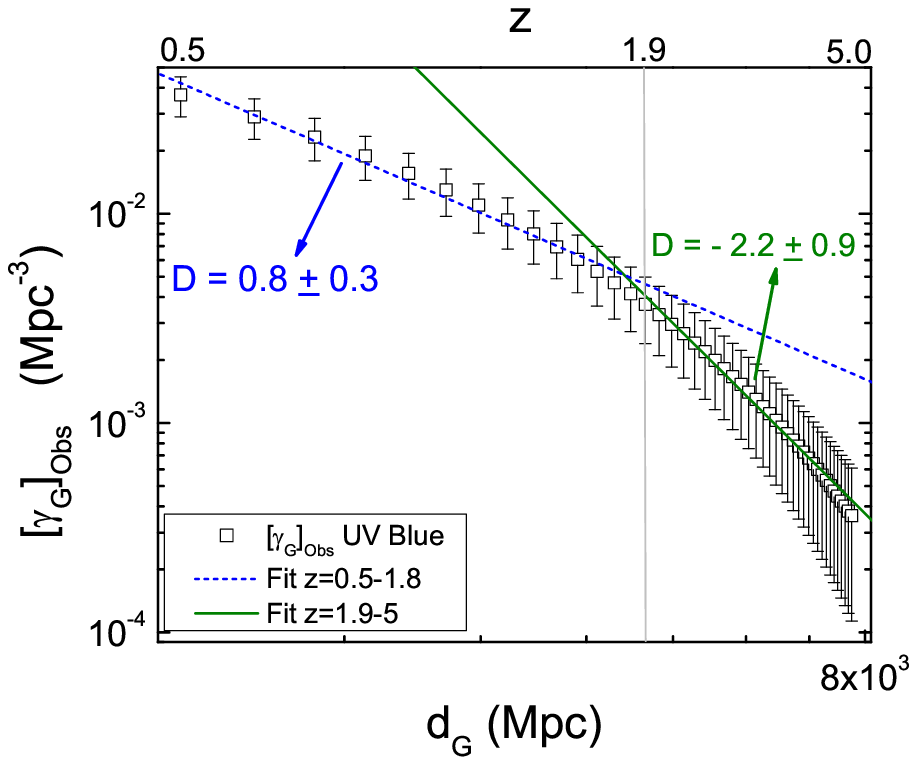} &
\includegraphics[width=5.3cm]{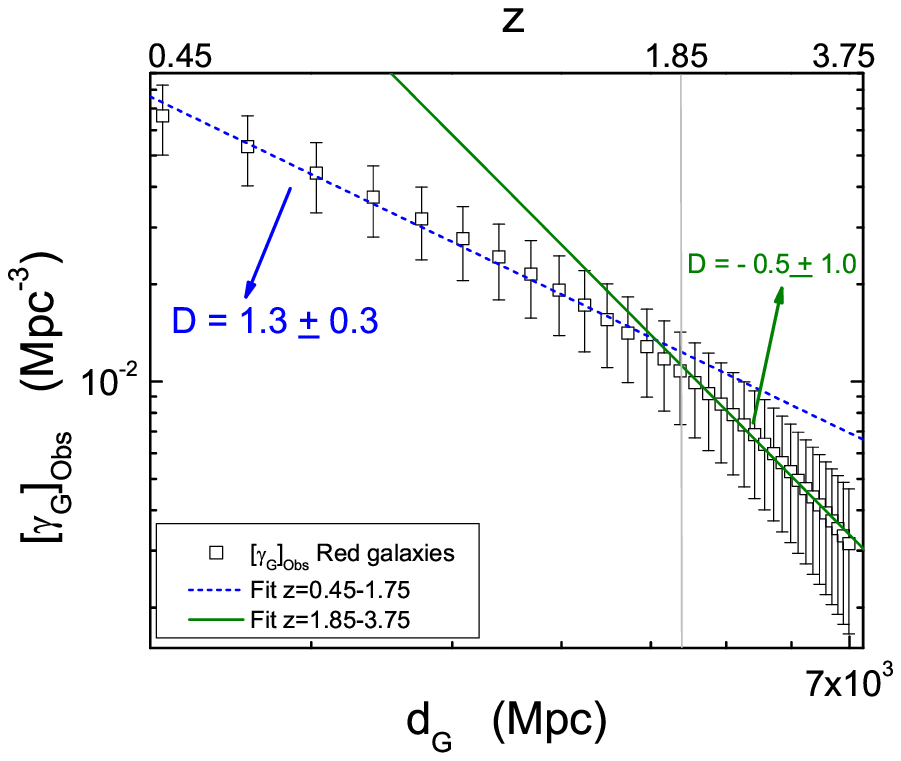} \\
\includegraphics[width=5.3cm]{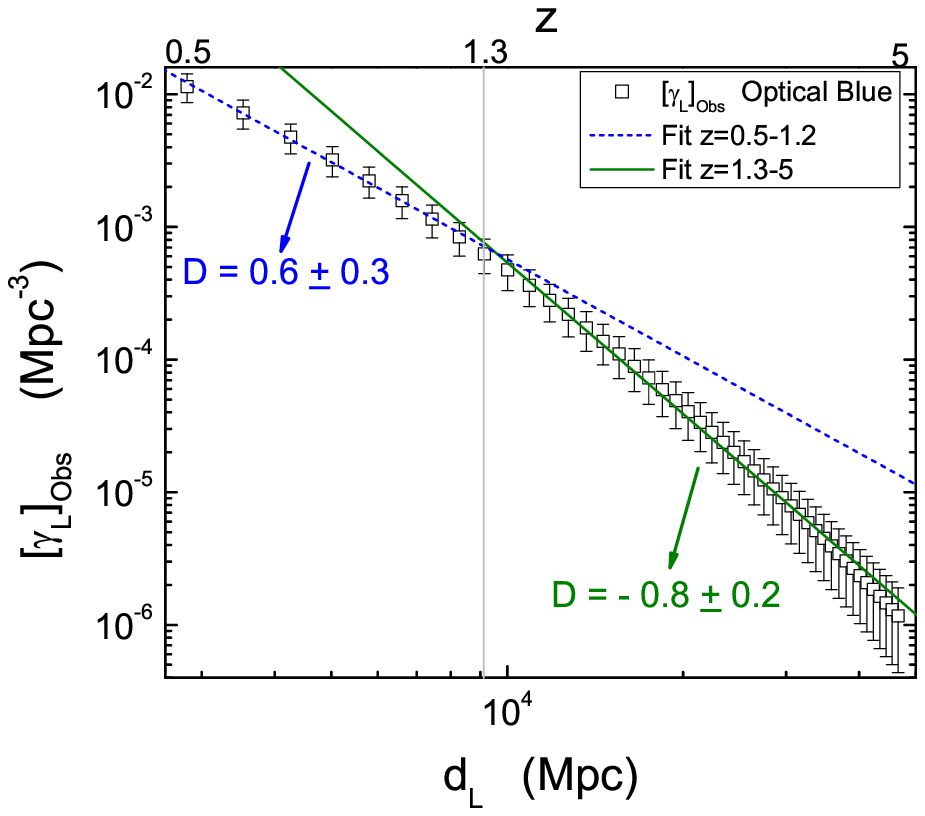} &
\includegraphics[width=5.3cm]{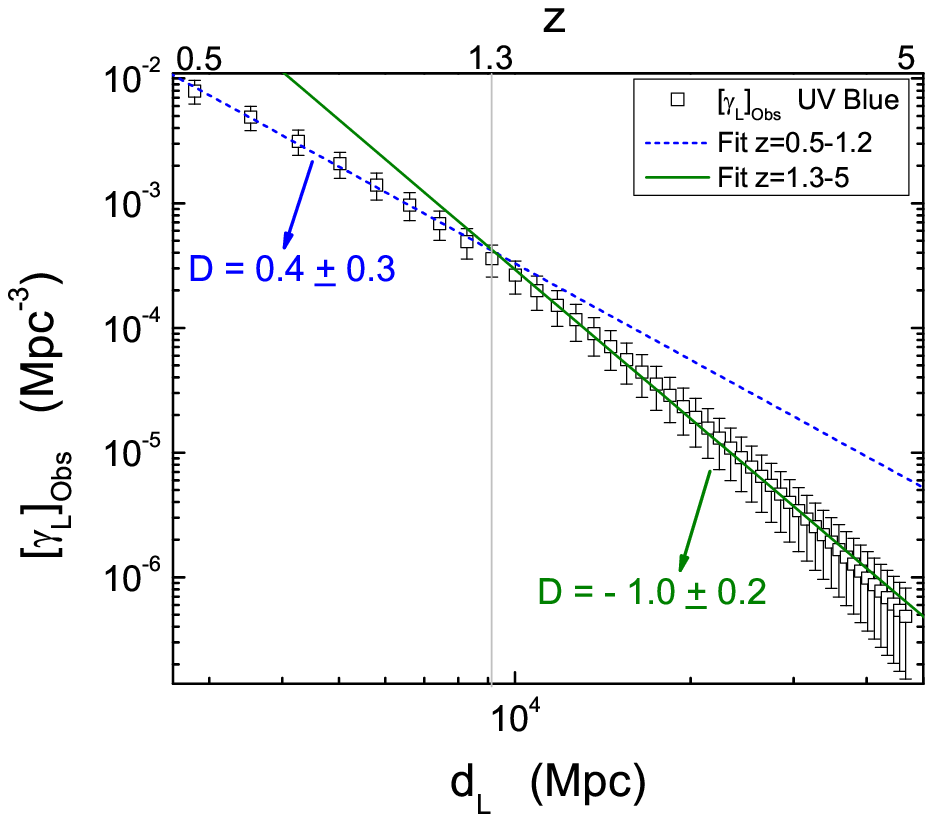} &
\includegraphics[width=5.3cm]{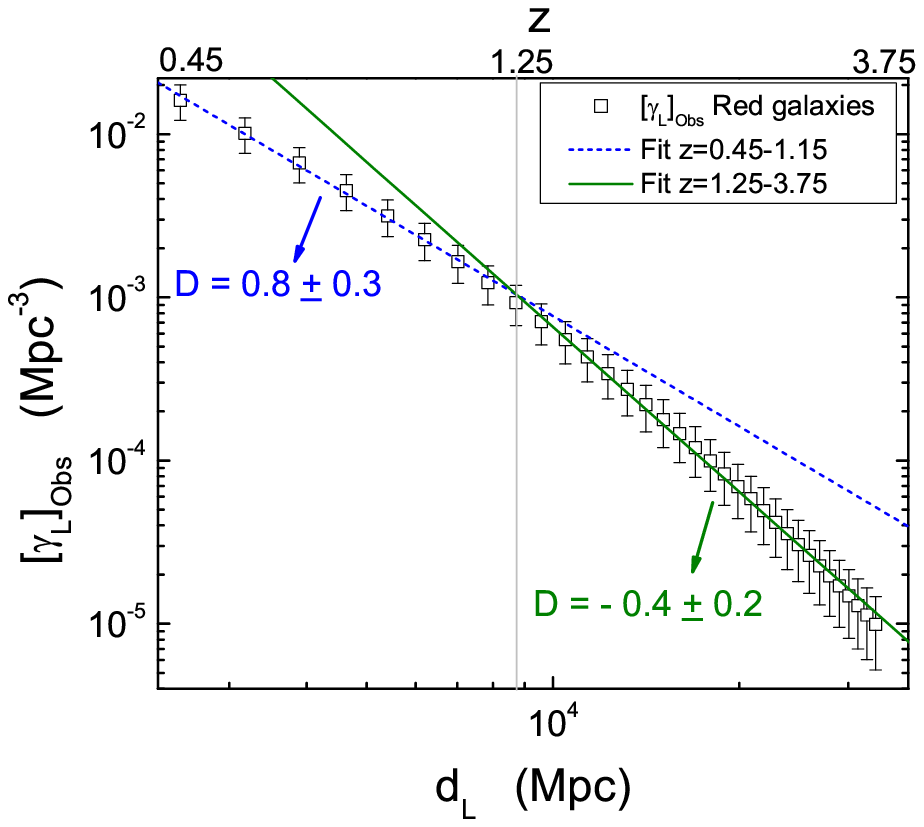} \\
\includegraphics[width=5.3cm]{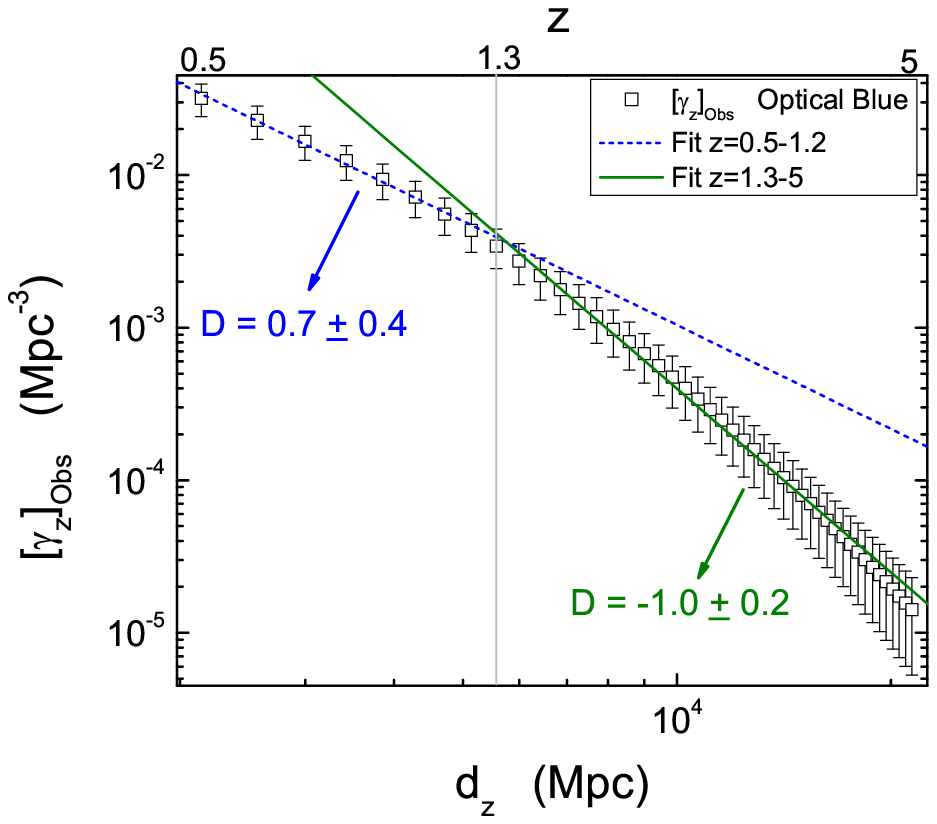} &
\includegraphics[width=5.3cm]{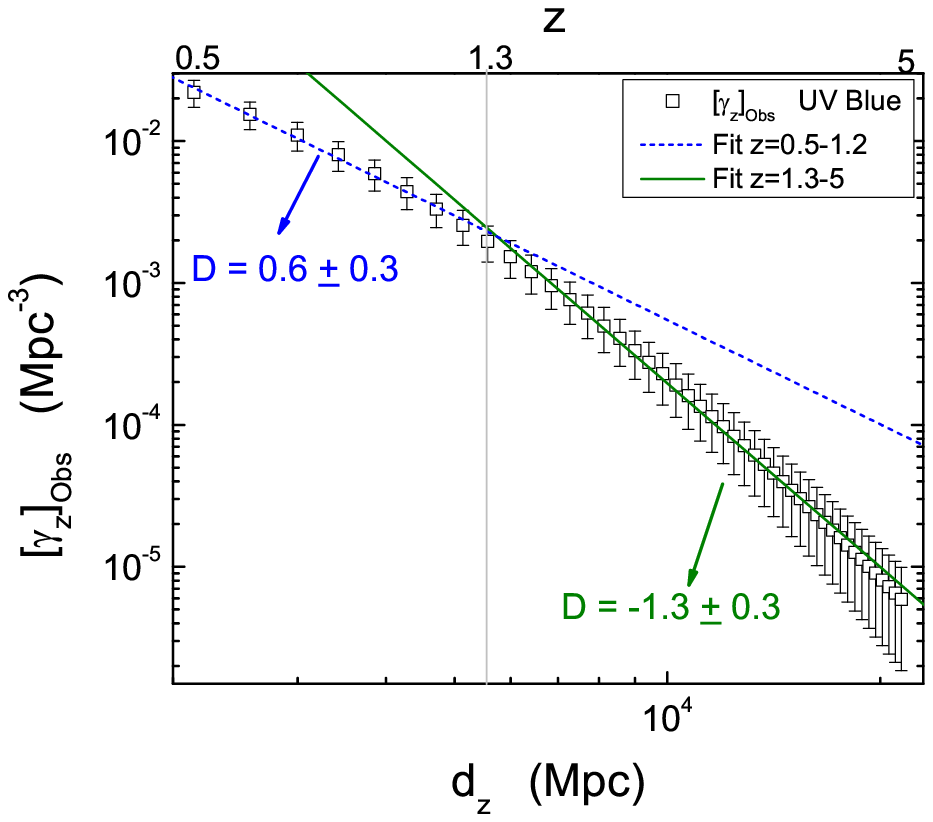} &
\includegraphics[width=5.3cm]{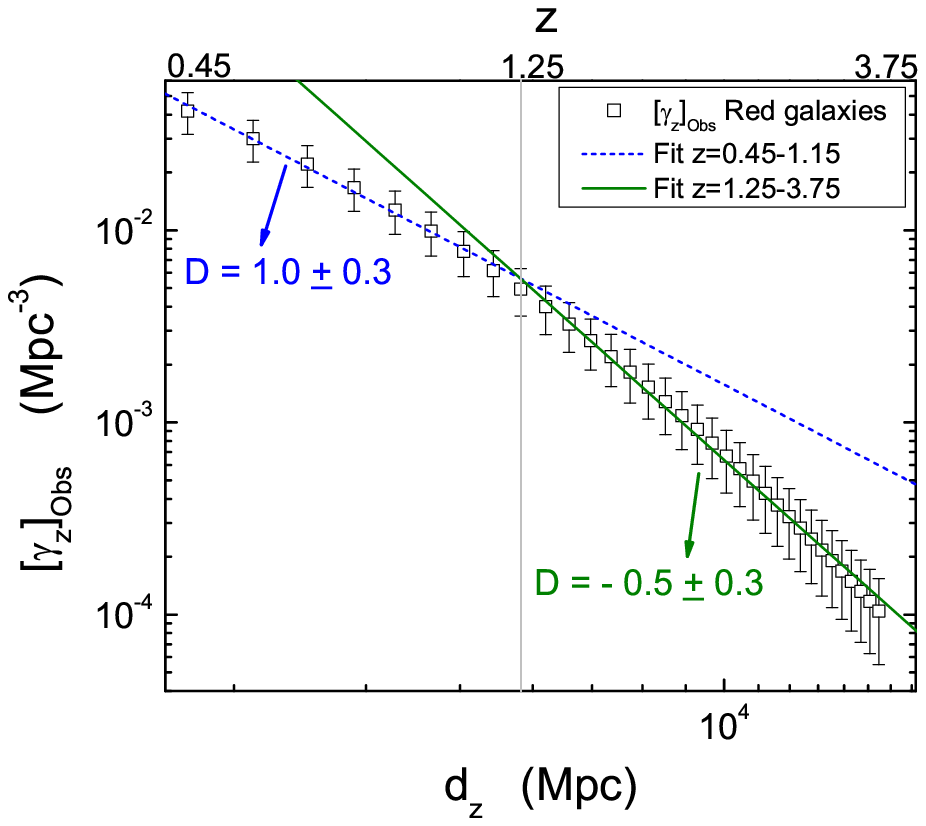} 
\end{array}$
\end{center}
\caption{Graphs of $\obs{\gamma_i} \times d_i$. $D$ is obtained by
a double linear fitting according to Eq.\ (\ref{gama3}).}
\lb{gamas}
\end{figure}

\begin{figure}
\begin{center}$
\begin{array}{ccc}
\includegraphics*[width=5.3cm]{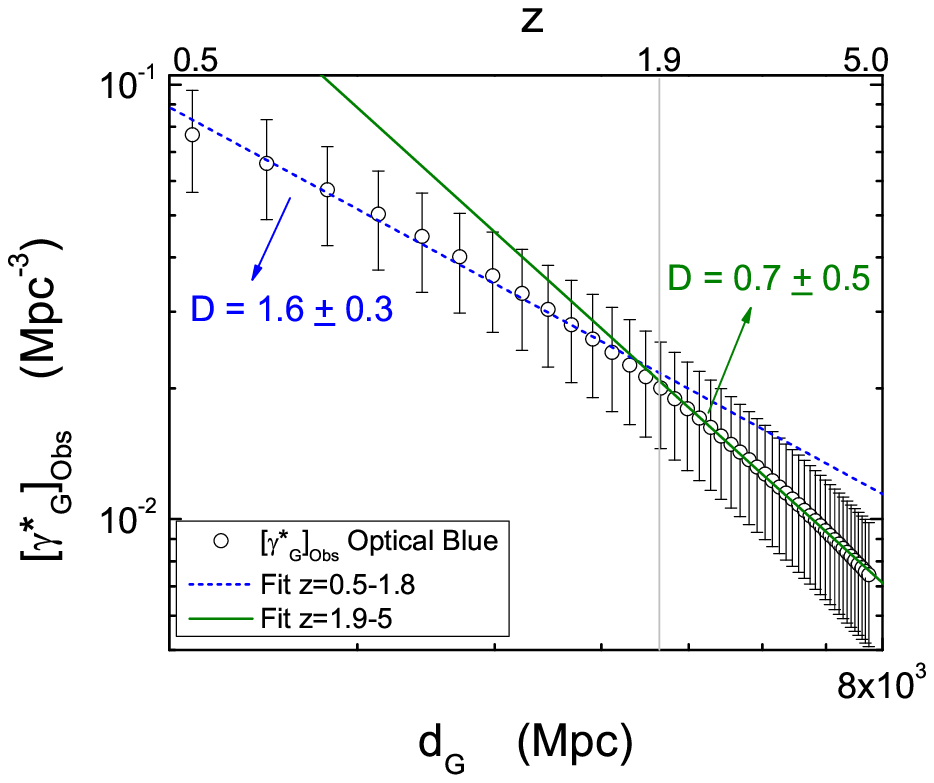} &
\includegraphics*[width=5.3cm]{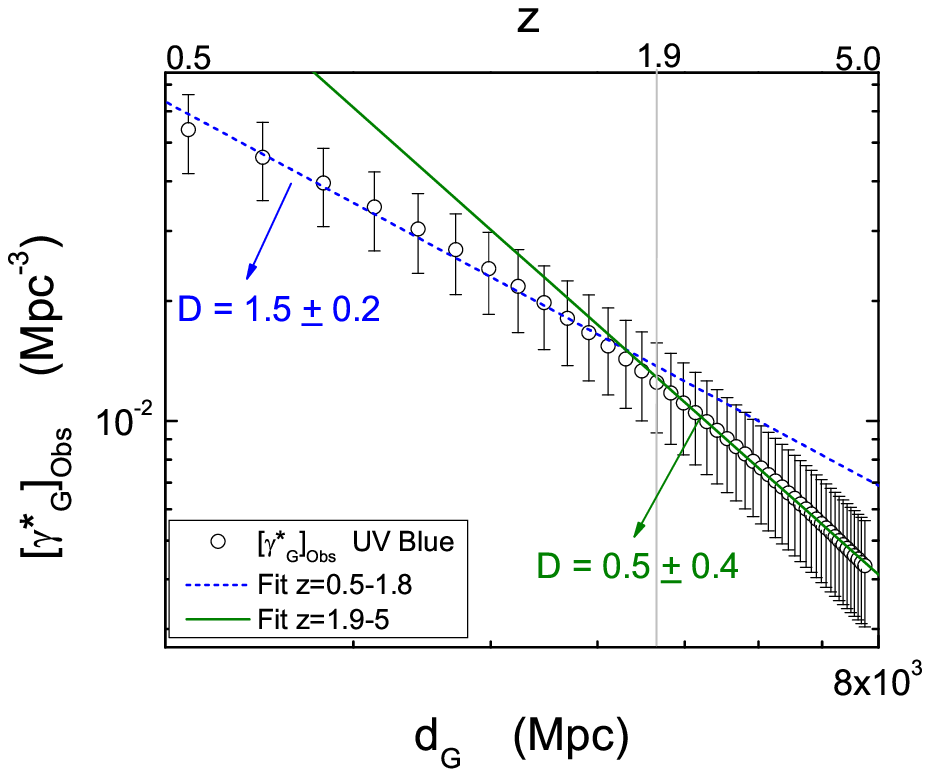} &
\includegraphics*[width=5.3cm]{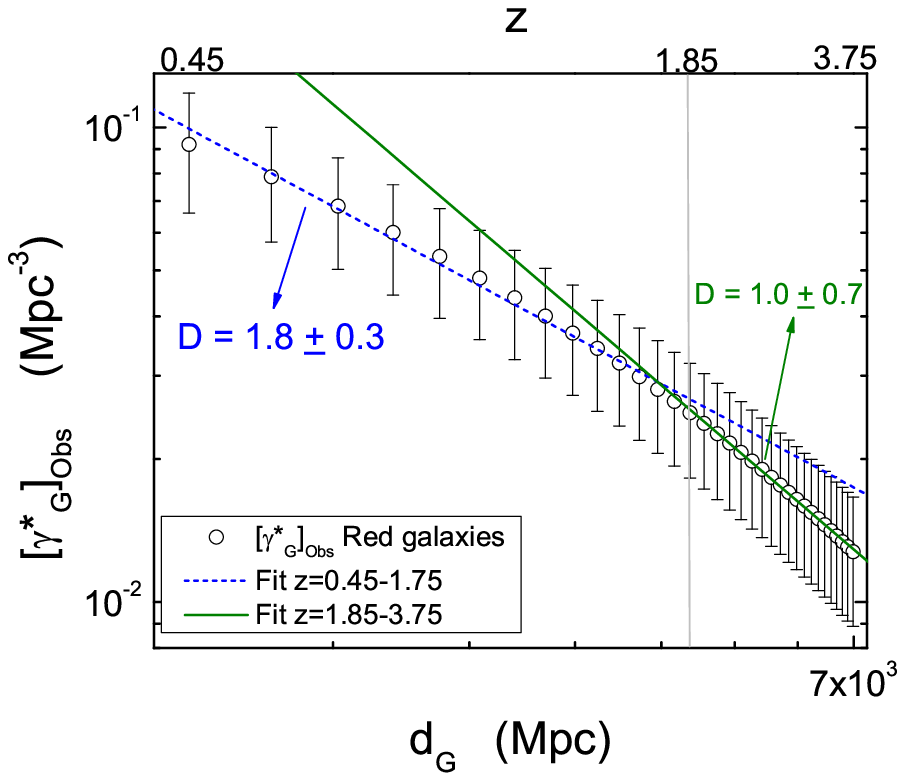} \\
\includegraphics*[width=5.3cm]{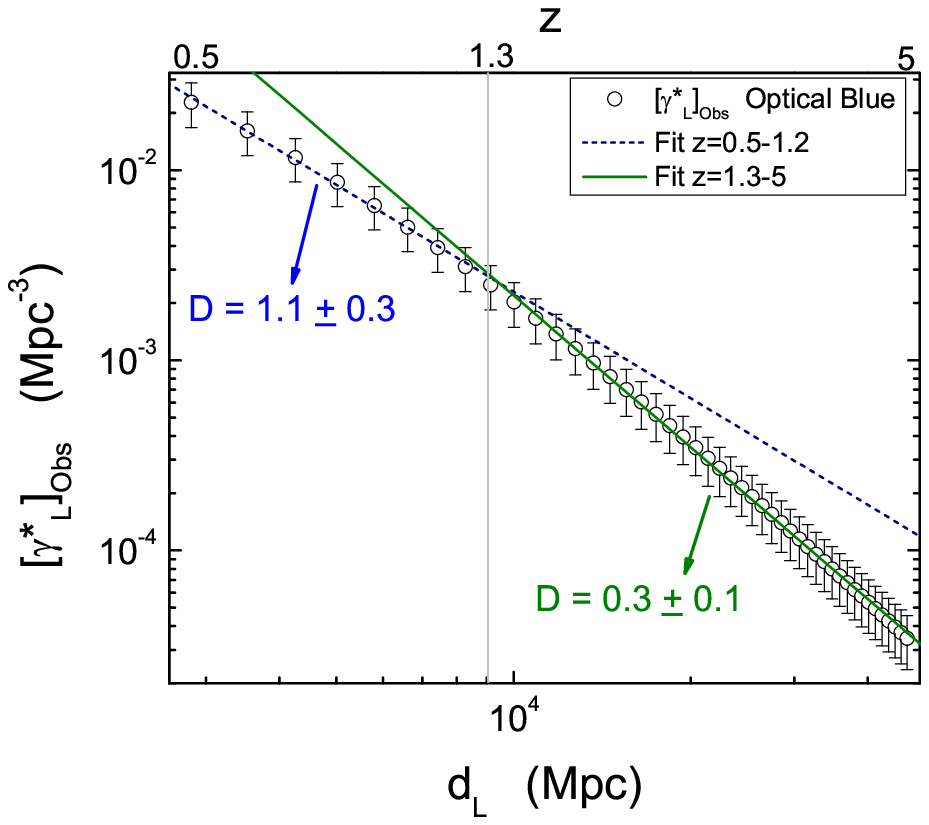} &
\includegraphics*[width=5.3cm]{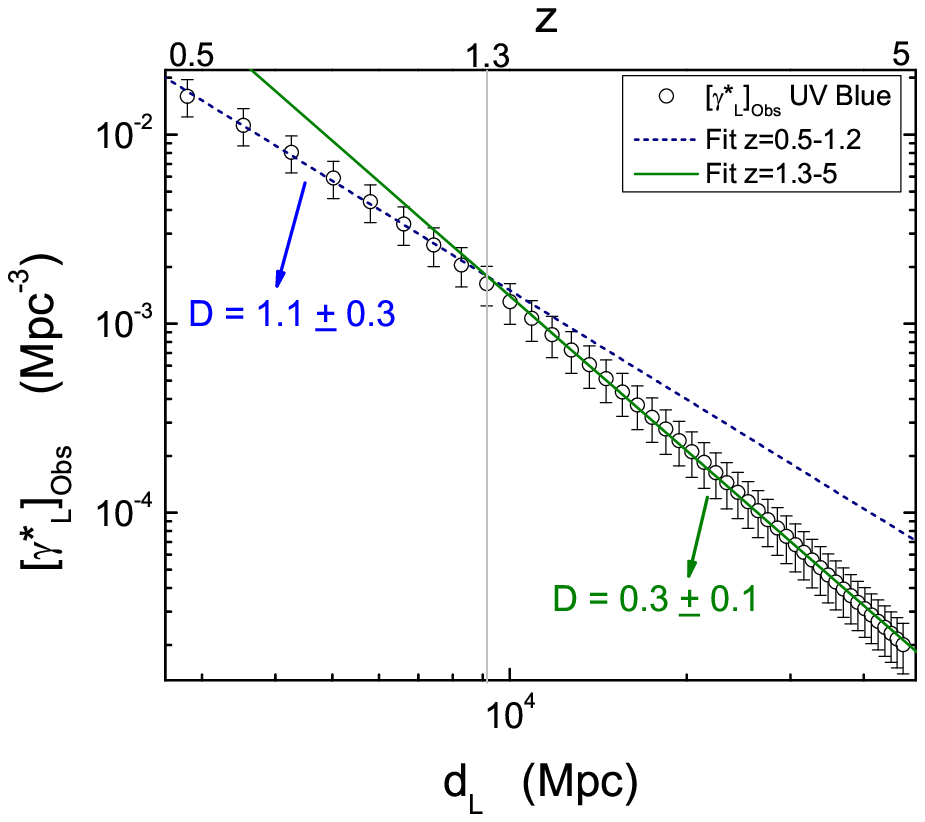} &
\includegraphics*[width=5.3cm]{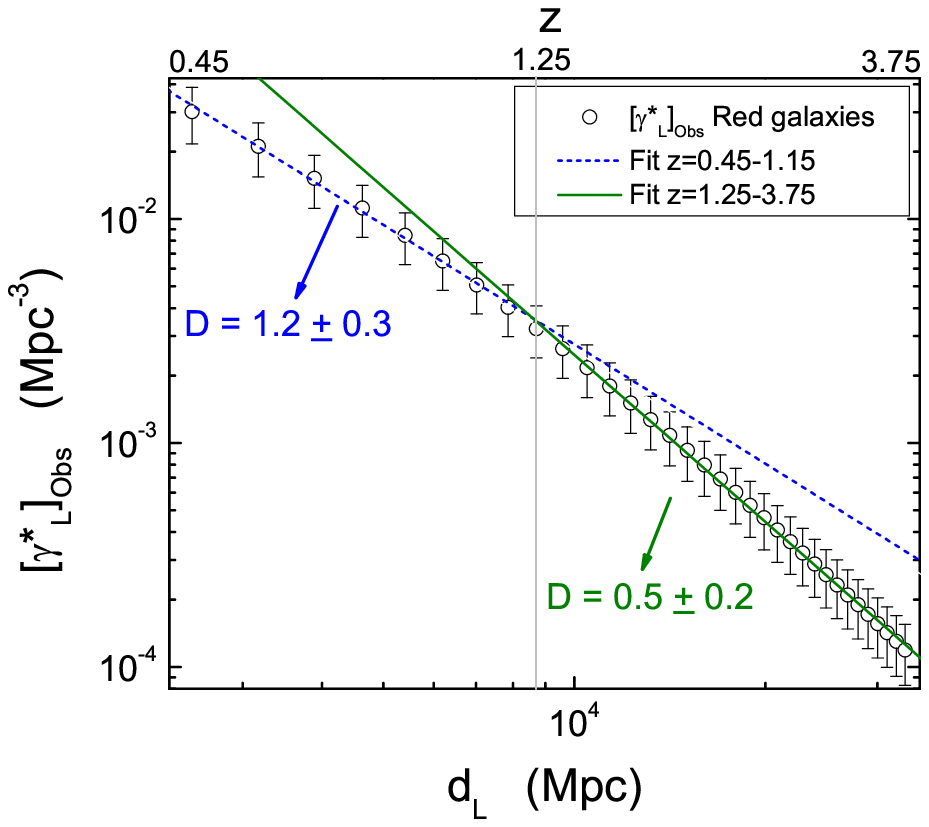} \\
\includegraphics*[width=5.3cm]{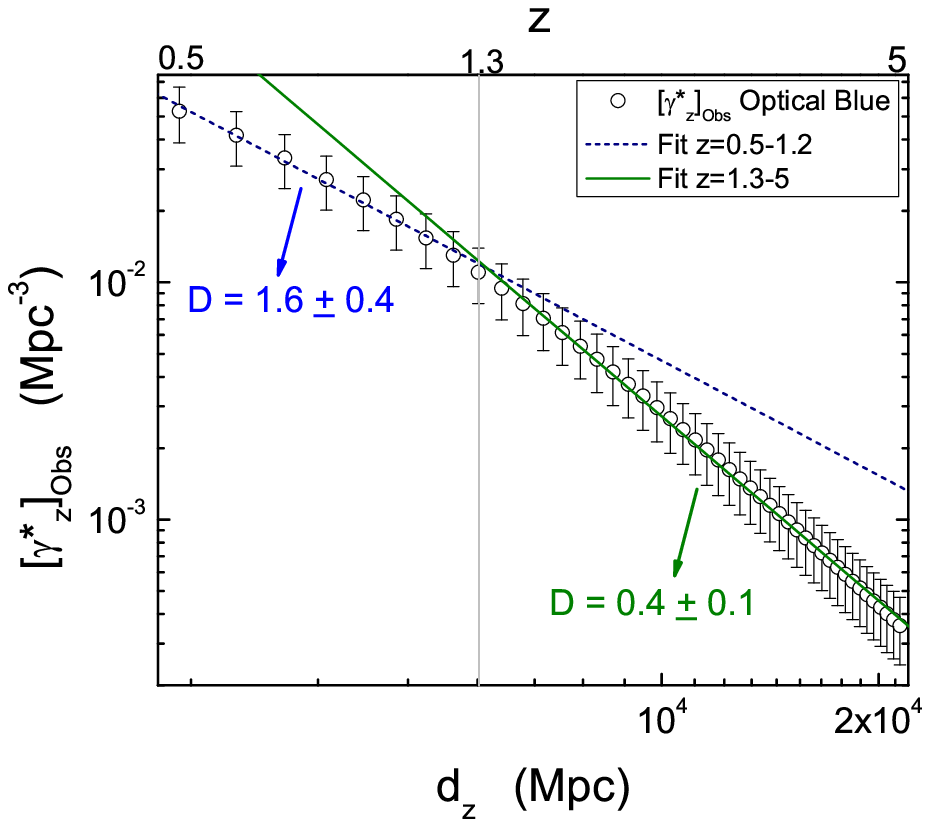} &
\includegraphics*[width=5.3cm]{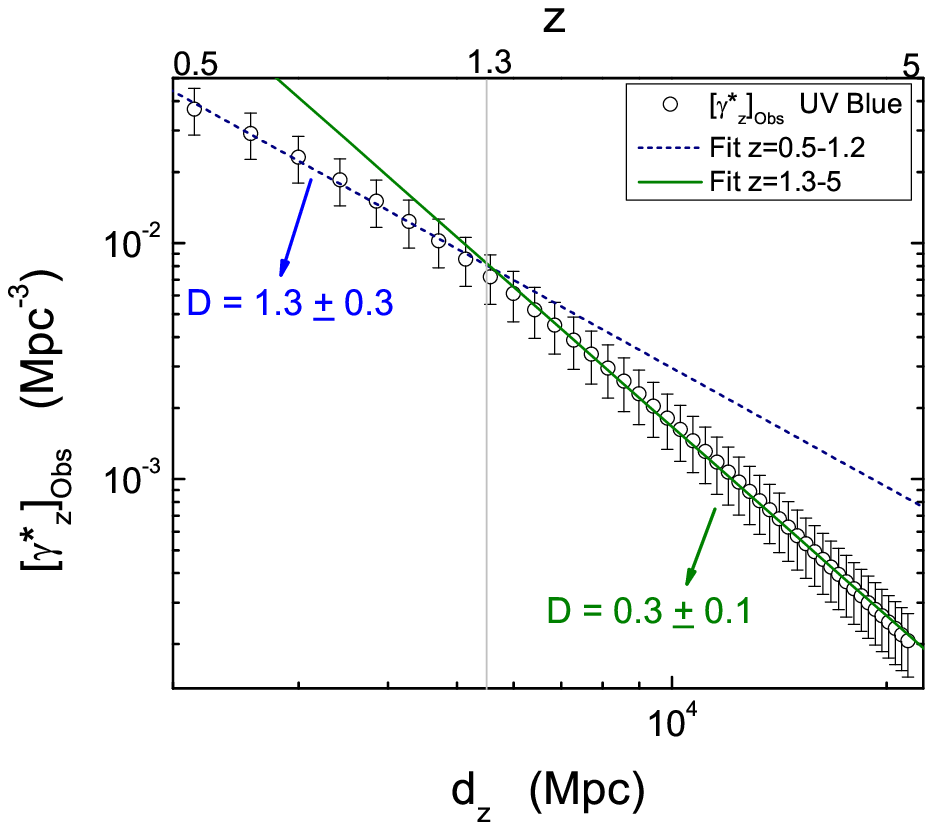} &
\includegraphics*[width=5.3cm]{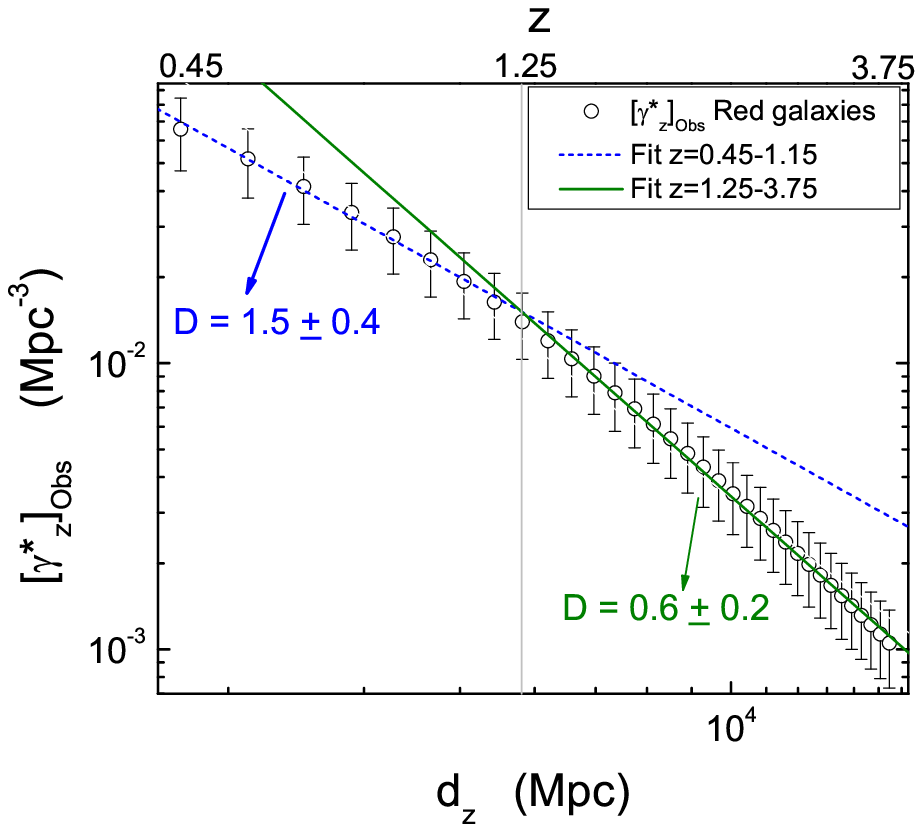} 
\end{array}$
\end{center}
\caption{Graphs of $\obs{\gamma_i^\ast} \times d_i$. $D$ is obtained by
a double linear fitting according to Eq.\ (\ref{gstar3}).}
\lb{gamasstar}
\end{figure}

\end{document}